\documentclass[sigconf, nonacm]{acmart}

\usepackage{pvldb}%

\renewcommand\vldbdoi{XX.XX/XXX.XX}
\renewcommand\vldbpages{XXX-XXX}
\renewcommand\vldbavailabilityurl{https://github.com/Inverse0/validation-aware-decomposition} 

\usepackage{CJKutf8}
\usepackage{xcolor}
\usepackage{graphicx}
\usepackage{enumitem}
\usepackage{listings}
\usepackage{tabularx}
\usepackage{booktabs}
\usepackage{colortbl}
\usepackage{multirow}
\usepackage{algorithm}
\usepackage{algpseudocode}
\usepackage{xspace}
\usepackage{tikz}
\usetikzlibrary{positioning,arrows.meta,fit,calc,shapes.misc}

\newcommand{\cut}[1]{}

\definecolor{todoblue}{RGB}{25,75,165}

\newcommand{\sys}{\textsc{DecoVal}\xspace}
\newcommand{\whole}{\textsc{Whole}\xspace}
\newcommand{\wholek}{\textsc{Whole}~$\times k$\xspace}
\newcommand{\syntactic}{\textsc{Syntactic}\xspace}
\newcommand{\randomk}{\textsc{Random}-$k$\xspace}
\newcommand{\rt}{\textsc{RT}\xspace}
\newcommand{\se}{\textsc{SE}\xspace}
\newcommand{\Dspace}{\mathcal{D}_{\mathcal{R}}(q)}
\newcommand{\classes}{\mathcal{C}}
\newcommand{\Compose}{\operatorname{Compose}}

\newcommand{\sqlkw}[1]{\texttt{#1}}

\lstdefinestyle{sqlstyle}{
    language=SQL,
    basicstyle=\ttfamily\footnotesize,
    keywordstyle=\bfseries,
    commentstyle=\itshape\color{gray},
    stringstyle=\color{blue},
    columns=fullflexible,
    keepspaces=true,
    breaklines=true,
    showstringspaces=false,
    tabsize=2,
    aboveskip=\smallskipamount,
    belowskip=\smallskipamount
}
\lstdefinestyle{blockstyle}{
    style=sqlstyle,
    basicstyle=\ttfamily\scriptsize,
    aboveskip=0pt,
    belowskip=0pt,
    xleftmargin=0pt,
    frame=none
}

\begin{document}
\title{Optimizing Query Decomposition for Reliable Validation}

\author{Shuo Zhang}
\affiliation{%
  \institution{Duke University}
  \city{Durham, North Carolina}
  \country{USA}
}
\email{shuo.zhang2@duke.edu}

\author{Sudeepa Roy}
\affiliation{%
  \institution{Duke University}
  \city{Durham, North Carolina}
  \country{USA}
}
\email{sudeepa@cs.duke.edu}

\author{Jun Yang}
\affiliation{%
  \institution{Duke University}
  \city{Durham, North Carolina}
  \country{USA}
}
\email{junyang@cs.duke.edu}

\begin{abstract}
Large language models (LLMs) have made querying databases more accessible, but they cannot guarantee that the SQL they generate is correct. Without a trusted reference query, a validator (a human or an LLM) must determine whether a generated query answers the intended question. Enterprise questions often require queries with multiple joins, nested subqueries, and several levels of aggregation, and validating such a query in a single judgment is error-prone. Decomposition can simplify this task by rewriting a query into smaller queries, or blocks, that recompose to the original, each checked against its own sub-question. 

Decompositions differ, however: smaller blocks make each judgment easier but require more judgments, and the best balance depends on the validator's ability. We formulate \emph{validation-aware query decomposition} as an optimization problem. 
Just as a query optimizer chooses among equivalent plans by estimated execution cost, we choose among decompositions by how reliably a given validator is expected to check them. 
We model how task difficulty and validator ability jointly affect validation reliability and define an objective that balances reliability against the number of required judgments. 
Over a space of decompositions derived from logical-plan rewrites, our search finds an optimal one.
We implement this approach in \sys{}, a validation pipeline that presents each block with its sub-question and input interface, combines local verdicts into a query-level decision, and localizes errors to individual blocks.

A user study shows that decomposition helps human validators catch and locate errors: with \sys{}, participants caught 91\% of the errors, up from 36\% without decomposition, and located the error in nearly every rejection. Experiments on BEAVER and Spider~2.0 with LLM validators show that \sys{} also increases error-detection recall over whole-query validation while maintaining high precision, and that the best granularity depends on the validator.

\end{abstract}

\maketitle

\vldbtopmatter

\section{Introduction}
\label{sec:introduction}

The rapid growth of AI-assisted querying is changing who can obtain value from data~\cite{floratou2024nl2sql}.
A user with limited SQL experience can now describe a complex business question in natural language and receive sophisticated SQL
within seconds from an LLM~\cite{gao2023texttosqlempoweredlargelanguage,katsogiannis2023survey,hong2025nextgenerationdatabaseinterfacessurvey}.
But how can a user know whether the generated SQL is correct?
A query can be syntactically valid and return plausible results but still return a semantically wrong answer; 
a user who trusts it acts on wrong information, which could lead to serious consequences~\cite{alrashed2026verification}.
The bottleneck has thus shifted rather than disappeared: the central question
is no longer only whether a model can produce a query, but whether the produced
query implements the intended analysis.

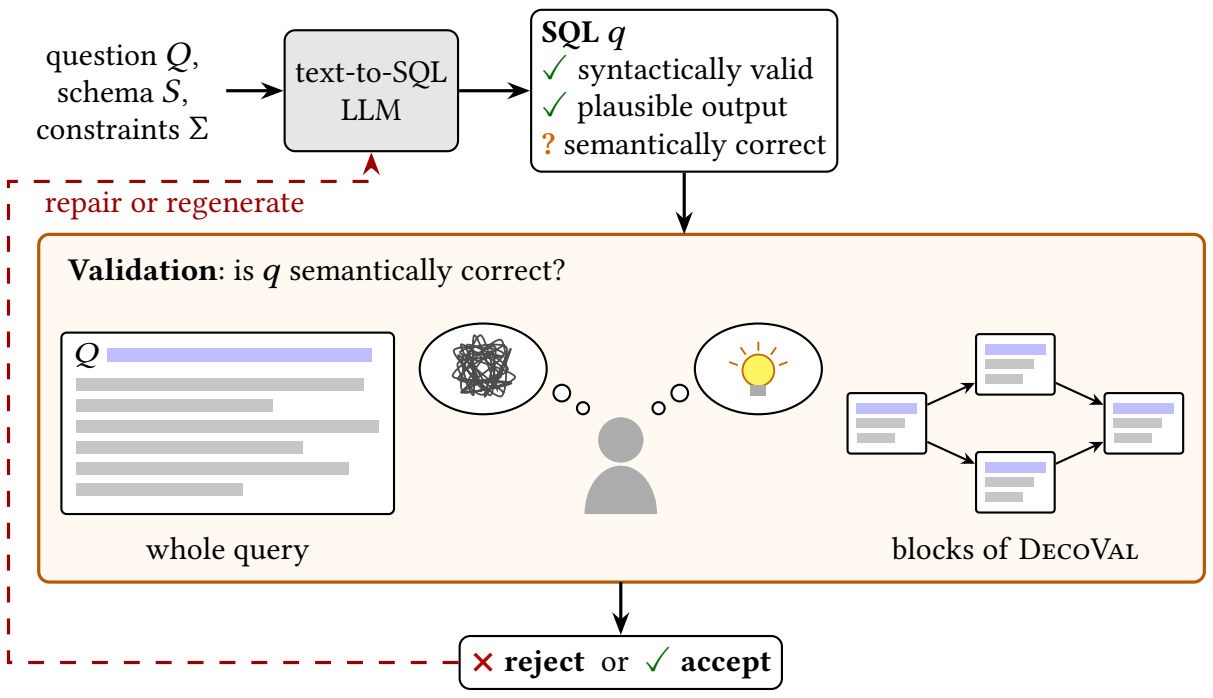
\begin{figure}[t]
\centering
\begin{tikzpicture}[
  x=1mm, y=1mm,
  font=\scriptsize,
  box/.style={draw, rounded corners=2pt, align=center, inner sep=2pt},
  arr/.style={-{Stealth[length=1.5mm]}, semithick},
  person/.pic={
    \fill[gray!65] (0,4.9) circle (1.5);
    \fill[gray!65] (-2.4,0) arc[start angle=180, end angle=0, x radius=2.4, y radius=3.2] -- cycle;
  },
  blk/.style={draw, fill=white, rounded corners=0.6pt, minimum width=5.2mm,
    minimum height=4mm, inner sep=0pt},
  flow/.style={-{Stealth[length=0.9mm, width=0.8mm]}, line width=0.4pt},
]
\node[align=center, anchor=east] (inp) at (15.9,27) {question $Q$,\\ schema $S$,\\ constraints $\Sigma$};
\node[box, fill=gray!20, minimum height=8mm] (gen) at (25.5,27) {text-to-SQL\\ LLM};
\node[box, fill=white, align=left, anchor=west] (qc) at (36,27)
  {\textbf{SQL $q$}\\
   {\color{green!40!black}\checkmark}~syntactically valid\\
   {\color{green!40!black}\checkmark}~plausible output\\
   {\color{orange!80!black}\textbf{?}}~semantically correct};
\draw[arr] (inp) -- (gen);
\draw[arr] (gen) -- (qc);
\draw[rounded corners=2pt, draw=orange!70!black, fill=orange!5, line width=0.6pt]
  (3.5,-5.5) rectangle (80.6,17.5);
\node[anchor=north west] at (4.3,17.1) {\textbf{Validation}: is $q$ semantically correct?};
\draw[arr] (qc.south) -- (qc.south |- 0,17.5);
\draw[rounded corners=1pt, fill=white] (5,-1) rectangle (27,11);
\node[font=\tiny\bfseries, inner sep=0pt] at (6.7,9.5) {$Q$};
\fill[blue!25] (8,9.05) rectangle (25.5,9.95);
\foreach \y/\l in {7.6/19, 6.2/13, 4.8/20, 3.4/15, 2.0/18, 0.6/11}
  \fill[gray!45] (6,\y-0.4) rectangle (6+\l,\y+0.4);
\foreach \n/\x/\y in {b1/59.6/5.0, b2/68.1/8.9, b3/68.1/1.1, b4/76.6/5.0} {
  \node[blk] (\n) at (\x,\y) {};
  \fill[blue!25] (\x-2.0,\y+0.65) rectangle (\x+2.0,\y+1.35);
  \fill[gray!45] (\x-2.0,\y-0.3) rectangle (\x+1.2,\y+0.3);
  \fill[gray!45] (\x-2.0,\y-1.25) rectangle (\x+0.5,\y-0.65);
}
\draw[flow] (b1) -- (b2);
\draw[flow] (b1) -- (b3);
\draw[flow] (b2) -- (b4);
\draw[flow] (b3) -- (b4);
\node[anchor=north] at (16,-1.4) {whole query};
\node[anchor=north] at (68.1,-1.4) {blocks of \sys};
\pic at (42,-1) {person};
\draw[fill=white] (39.5,6.0) circle (0.4);
\draw[fill=white] (38.1,7.0) circle (0.55);
\draw[fill=white] (32.9,8.6) ellipse [x radius=4.2, y radius=3.0];
\draw[black!70, line width=0.35pt, shift={(32.9,8.6)}] plot[smooth, tension=0.6] coordinates
  {(1.42,1.60) (-1.65,-0.66) (1.86,-0.94) (-1.46,-1.55) (1.27,-0.07) (-0.93,1.18) (0.54,-1.18) (1.81,1.32) (-0.02,-1.70) (-1.68,0.37) (1.27,0.81) (-1.33,0.83) (-0.64,-1.72) (-0.58,2.22) (0.91,-1.33) (-0.19,2.23) (-1.40,-1.37) (2.11,-0.54) (-1.53,1.08) (1.54,1.41) (0.11,-1.83) (0.76,1.47) (-2.01,-0.48) (1.53,-0.89) (-1.34,-1.48) (1.34,1.00) (-1.95,1.19) (1.16,-1.82) (1.17,1.92) (-1.46,-0.51) (1.17,-0.48) (-1.47,1.61) (0.44,-1.75) (0.59,1.18) (-1.05,-1.03) (1.63,-0.01) (-1.07,1.58) (-0.07,-1.29) (-1.63,0.91) (1.37,-1.58)};
\draw[fill=white] (44.5,6.0) circle (0.4);
\draw[fill=white] (45.9,7.0) circle (0.55);
\draw[fill=white] (51.1,8.6) ellipse [x radius=4.2, y radius=3.0];
\begin{scope}[shift={(51.1,8.35)}]
  \foreach \a in {20,55,90,125,160}
    \draw[orange!80!black, line width=0.35pt] (\a:1.45) -- (\a:2.05);
  \fill[yellow!75, draw=orange!80!black, line width=0.35pt] (0,0.15) circle (1.05);
  \fill[gray!65] (-0.5,-1.55) rectangle (0.5,-0.85);
\end{scope}
\node[box, fill=white, anchor=north] (out) at (42,-9)
  {{\color{red!70!black}$\boldsymbol{\times}$}~\textbf{reject}\enspace or\enspace
   {\color{green!40!black}\checkmark}~\textbf{accept}};
\draw[arr] (42,-5.5) -- (out.north);
\draw[arr, dashed, red!60!black] (out.west) -- (1.5,0 |- out.west) -- (1.5,21) -| (gen.south);
\node[anchor=north, text=red!60!black, inner sep=1pt] at (12.5,20.8) {repair or regenerate};
\end{tikzpicture}
\caption{Validating LLM-generated SQL. The
generated query $q$ can be valid and return plausible output, but a validator needs to decide whether it is semantically correct. 
}
\Description{A text-to-SQL LLM turns a question, a schema, and its constraints into a SQL query
that is marked syntactically valid and with plausible output, while its semantic
correctness carries a question mark. The query enters a validation step. Inside it, a validator looks
at the question and the whole query on the left and thinks a tangle of lines,
or at the blocks of DecoVal on the right, four small blocks connected as a DAG,
each with a requirement and a short SQL query, and thinks of a light bulb. The
step ends in one box, reject or accept; a dashed arrow from reject back to the
LLM marks repairing or regenerating the query.}
\label{fig:teaser}
\end{figure}

In this work, we address the problem of \emph{Reference-Free Validation} (Figure~\ref{fig:teaser}): 
given a natural-language question~$Q$, a database schema~$S$ with constraints~$\Sigma$ (e.g., keys and foreign keys), and a candidate SQL
query~$q$ over~$S$, we want to determine whether $q$ correctly answers $Q$.
When a trusted reference implementation query for $Q$ exists, assurance reduces to equivalence checking, for which provers
and bounded checkers are available~\cite{wang2024qed,he2024verieql}.
In practice, such a reference rarely exists: writing it is the very task
the user delegated to the LLM. Executing the candidate cannot settle the
question: enterprise data is often private~\cite{chen2026beaver}, and an
execution only allows checking on one database instance,
where subtle errors, such as mishandling duplicates, \sqlkw{NULL}s, or missing joins, may not surface.
Without a reference, the specification is the request itself, and the decision rests
with a \emph{validator}---a human user or an LLM---who reads the candidate and judges 
whether it answers the question.

\begin{figure}[t]
\centering
\begin{lstlisting}[style=sqlstyle, basicstyle=\ttfamily\fontsize{3.5}{3.1}\selectfont, keywordstyle={\color[RGB]{0,110,150}}, stringstyle={\color[RGB]{190,90,0}}, morekeywords={ROLLUP,GROUPING,OVER,PARTITION,STDDEV,BIGINT,TEXT,PRECISION}, deletekeywords={ACTION}, breakatwhitespace=true, breakindent=0pt, aboveskip=0pt, belowskip=0pt]
WITH non_shared_network_actions AS (SELECT DISTINCT nr.action, n.id AS network_id FROM networkrbacs AS nr JOIN networks AS n ON n.id = nr.object_id WHERE n.id NOT IN (SELECT networkrbacs.object_id FROM networkrbacs WHERE networkrbacs.action = 'access_as_shared' AND (networkrbacs.target_tenant = '*' OR networkrbacs.target_tenant = '9f13bf11e06e4d81b44c8c3e62d4362b'))), action_network_stats AS (SELECT DISTINCT nna.action, nna.network_id, COUNT(s.id) OVER (PARTITION BY nna.action, nna.network_id) AS subnet_count, STDDEV(s.ip_version) OVER (PARTITION BY nna.action, nna.network_id) AS network_stddev_ip_version FROM non_shared_network_actions AS nna JOIN subnets AS s ON s.network_id = nna.network_id WHERE s.enable_dhcp = 1), qualifying_actions AS (SELECT action FROM action_network_stats GROUP BY action HAVING SUM(subnet_count) > 5), qualifying_network_stats AS (SELECT ans.* FROM action_network_stats AS ans JOIN qualifying_actions AS qa ON qa.action = ans.action), action_rollup AS (SELECT CASE WHEN GROUPING(action) = 1 THEN 'GRAND TOTAL' ELSE 'ACTION SUBTOTAL' END AS row_kind, CASE WHEN GROUPING(action) = 1 THEN NULL ELSE action END AS rbac_action, AVG(network_stddev_ip_version) AS avg_stddev_ip_version, SUM(subnet_count)::bigint AS subnet_count, COUNT(*)::bigint AS network_count FROM qualifying_network_stats GROUP BY ROLLUP(action)), active_shared_networks AS (SELECT n.id AS network_id, n.status AS network_status FROM networks AS n WHERE n.status = 'ACTIVE' AND EXISTS (SELECT 1 FROM networkrbacs AS nr WHERE nr.object_id = n.id AND nr.action = 'access_as_shared')), network_subnet_counts AS (SELECT DISTINCT asn.network_id, asn.network_status, COUNT(s.id) OVER (PARTITION BY asn.network_id)::bigint AS subnet_count FROM active_shared_networks AS asn LEFT JOIN subnets AS s ON s.network_id = asn.network_id), classified_networks AS (SELECT nsc.network_id, nsc.network_status, nsc.subnet_count, CASE WHEN nsc.subnet_count > AVG(nsc.subnet_count) OVER () + STDDEV(nsc.subnet_count) OVER () THEN 'High Subnet Count' ELSE 'Normal Subnet Count' END AS subnet_count_classification FROM network_subnet_counts AS nsc), network_category_rollup AS (SELECT DISTINCT 'CATEGORY TOTAL'::text AS row_kind, network_status, subnet_count_classification, COUNT(*) OVER (PARTITION BY network_status, subnet_count_classification)::bigint AS network_count FROM classified_networks), network_status_rollup AS (SELECT DISTINCT 'STATUS SUBTOTAL'::text AS row_kind, network_status, NULL::text AS subnet_count_classification, COUNT(*) OVER (PARTITION BY network_status)::bigint AS network_count FROM classified_networks), network_grand_rollup AS (SELECT 'GRAND TOTAL'::text AS row_kind, NULL::text AS network_status, NULL::text AS subnet_count_classification, COUNT(*)::bigint AS network_count FROM classified_networks), network_rollup AS (SELECT row_kind, network_status, subnet_count_classification, network_count FROM network_category_rollup UNION ALL SELECT row_kind, network_status, subnet_count_classification, network_count FROM network_status_rollup UNION ALL SELECT row_kind, network_status, subnet_count_classification, network_count FROM network_grand_rollup), combined AS (SELECT 'RBAC_ACTION_STDDEV'::text AS report_section, ar.row_kind, ar.rbac_action, NULL::text AS network_id, NULL::text AS network_status, NULL::text AS subnet_count_classification, ar.avg_stddev_ip_version, ar.subnet_count, ar.network_count FROM action_rollup AS ar UNION ALL SELECT 'ACTIVE_SHARED_NETWORKS', 'NETWORK', 'access_as_shared', cn.network_id, cn.network_status, cn.subnet_count_classification, NULL::double precision, cn.subnet_count, 1::bigint FROM classified_networks AS cn UNION ALL SELECT 'ACTIVE_SHARED_NETWORKS', nr.row_kind, 'access_as_shared', NULL::text, nr.network_status, nr.subnet_count_classification, NULL::double precision, NULL::bigint, nr.network_count FROM network_rollup AS nr) SELECT report_section, row_kind, rbac_action, network_id, network_status, subnet_count_classification, avg_stddev_ip_version, subnet_count, network_count FROM combined ORDER BY CASE report_section WHEN 'RBAC_ACTION_STDDEV' THEN 1 ELSE 2 END, CASE row_kind WHEN 'ACTION SUBTOTAL' THEN 1 WHEN 'NETWORK' THEN 1 WHEN 'CATEGORY TOTAL' THEN 2 WHEN 'STATUS SUBTOTAL' THEN 3 WHEN 'GRAND TOTAL' THEN 4 ELSE 5 END, rbac_action, network_status, subnet_count_classification, network_id
\end{lstlisting}
\caption{A candidate 
for BEAVER question neutron\_152 \cite{chen2026beaver} with 13 CTEs, 20 \sqlkw{SELECT}s; shown for its size, not to be read.
}
\Description{An SQL query of about 4,400 characters with thirteen common table
expressions, window aggregates, a rollup, unions, and NOT IN and EXISTS
subqueries, printed as a dense block of text.}
\label{fig:beaver}
\end{figure}

One reason why this judgment is hard is that it is monolithic: the validator must
reason with the whole query including every join, filter, aggregation, nested structure, and correlation at once.
Figure~\ref{fig:beaver} shows a query that an LLM generated for a
question of the BEAVER enterprise benchmark (neutron\_152, \cite{chen2026beaver}). 
It chains
thirteen CTEs and combines window aggregates, a rollup, unions, and
\sqlkw{NOT} \sqlkw{IN} and \sqlkw{EXISTS} subqueries. It runs and returns plausible
rows, yet it does not answer the question. Few validators can assess such a
query as one unit.
Studies of SQL users find that errors concentrate on nested and correlated subqueries,
self-joins, negation, and \textsc{NULL} semantics~\cite{ahadi2015,taipalus2018,toussaint2022},
and that users' ability to judge their own queries declines as queries grow~\cite{casterella2013}.
LLMs falter on the same constructs~\cite{ning2024,li2024dawn,zhao2025}.
Recent work supplies more evidence for this judgment, for example by
translating the query back to natural language or by executing it on synthetic
inputs~\cite{alrashed2026verification}, but the judgment still covers the
whole query.

Our key insight is that a query is much easier to validate when it is simple for the
validator, that is, when the validator can understand it easily.
We therefore \emph{decompose} the candidate query into smaller
blocks tailored to the validator (shown as ``blocks of \sys'' in Figure~\ref{fig:teaser}). Each block is a SQL query that isolates one
local validation task: the validator checks it against its own sub-question,
using an interface that describes what each
input computes and the constraints it carries. Together, the blocks
recompose to exactly the candidate. Presenting a query as explained steps
already raises the accuracy with which people validate generated
SQL~\cite{tian2023}; decomposition makes those steps precise and checkable. 

\begin{figure*}[t]
\centering
\includegraphics[width=0.8\textwidth,trim=0 10 0 17,clip]{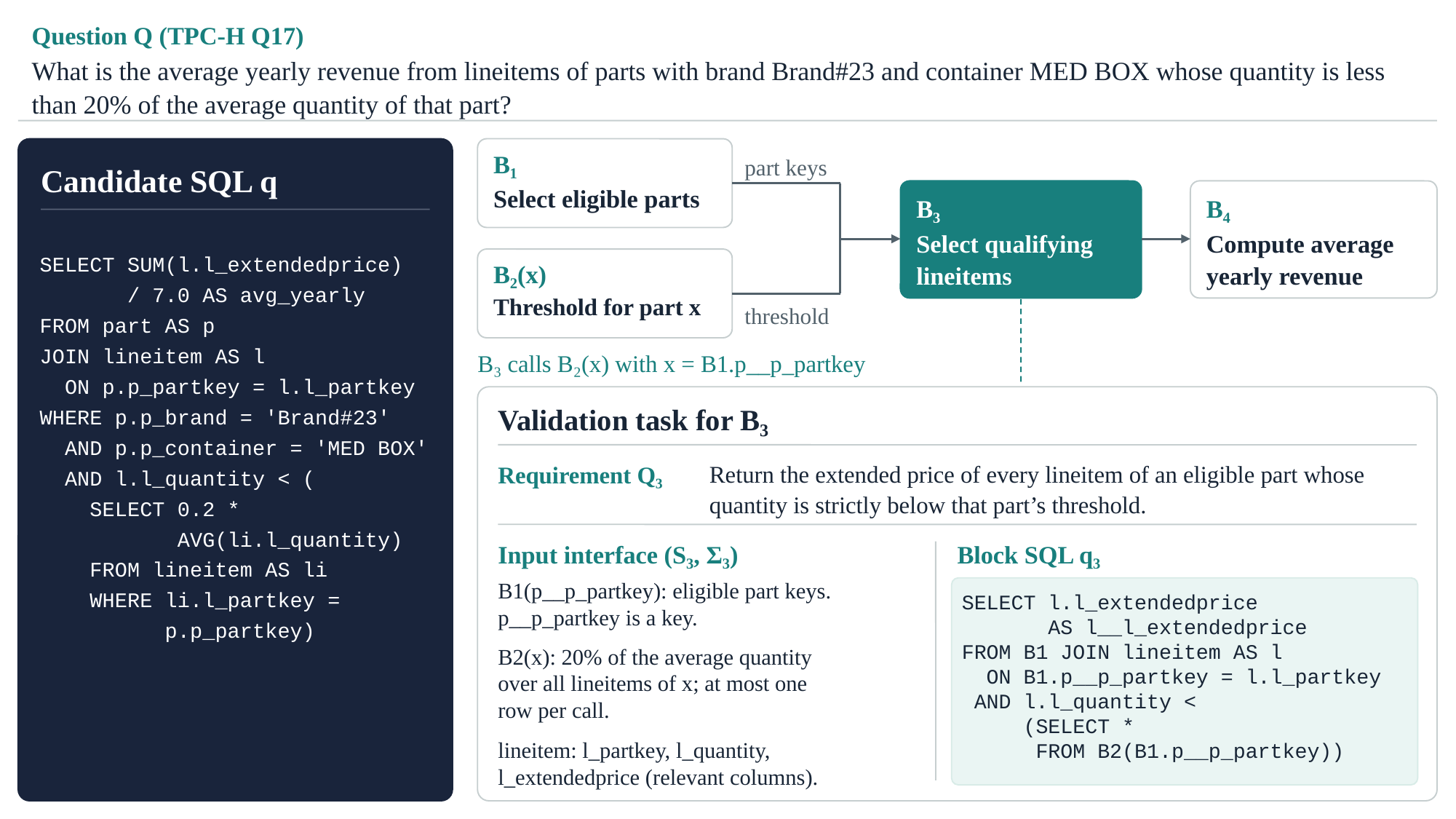}
\caption{Running example. Left: a candidate SQL query.
Right: a decomposition into four blocks and the validation task for B3.
B2 is parameterized by a part key and returns that part's quantity threshold;
B3 supplies the parameter when calling B2.}
\Description{The Q17 candidate is decomposed into four blocks. B1 selects
eligible parts, B2(x) computes a quantity threshold for part x, B3 uses both
to select qualifying lineitems, and B4 computes average yearly revenue.
The expanded B3 task shows its requirement, input descriptions and constraints,
and SQL with an explicit parameterized call to B2.}
\label{fig:q17}
\end{figure*}

Consider TPC-H Q17~\cite{tpch_spec}, which asks for the average yearly revenue
from lineitems whose quantity is below 20\% of the average quantity for the same
part, restricted to one brand and container. Judged as a whole, the candidate in
Figure~\ref{fig:q17} requires tracking part eligibility, a correlated average, a
per-part threshold, and a final aggregate simultaneously. The decomposition in
the same figure asks four local questions instead: which parts are eligible; what
is the threshold for a given part; which lineitems fall below their part's
threshold; and what is their total revenue per year.
Suppose a candidate omitted the correlation predicate
\texttt{li.l\_partkey = p.p\_partkey} and thus compared every lineitem with
20\% of the average quantity over \emph{all} parts. Judged as a whole, the
query still reads as a plausible threshold filter. Judged block by block, the
defect stands out: the threshold block should return a value for a given part,
yet its SQL takes no part as input.

Decomposition also introduces a fundamental tradeoff: smaller blocks reduce the load of each
judgment, but every boundary adds a judgment, adds interface context to
absorb, and risks separating information that a judgment needs. The best
decomposition therefore depends on the validator. A user fluent in correlated aggregates 
prefers to see a threshold subquery together with its consumer;
another finds it easier when the two are separate.

\smallskip
\noindent
\textbf{Challenges.}
Turning this insight into a solution raises four problems.
(1)~\textbf{Semantic preservation:} A decomposition must preserve the
candidate's semantics exactly, including row multiplicities and any errors, so
that judging its blocks amounts to judging the candidate. Relying on an LLM for decomposition can silently change what is being validated.
(2)~\textbf{Difficulty:} Choosing block boundaries requires a model of what
makes a validation task hard for a particular validator, from the SQL inside a
block and the inputs it reads to whether its output means something in terms
of the question. There is also a trade-off between the difficulty of each task
and the number of tasks.
(3)~\textbf{Search:} The space of decompositions grows with join orders, filter
placements, and subquery boundaries. Moreover, the same task can occur at
several places in a query
but only needs to be judged once,
which complicates local search decisions.
(4)~\textbf{Task assignment and composition:} Every block needs a requirement
that reflects the user's intent, and a
faithful description of what its inputs actually compute; the local verdicts must then
combine into a verdict on the whole query.

\medskip
\noindent
\textbf{Our Contributions.}~
In this paper, we introduce \sys{}, a system that addresses these challenges with
\emph{validation-aware query decomposition}: just as a query optimizer chooses among
equivalent plans by estimated execution cost, \sys{} chooses among
semantics-preserving decompositions by how reliably a given validator is
expected to judge them (Figure~\ref{fig:overview}). We make the following contributions in this paper:

\begin{enumerate}[leftmargin=*]
\item 
{\bf A formulation of validation-aware query decomposition:}
We leverage the idea of decomposition to tackle the Reference-Free Validation problem,
by formalizing validation-aware query decomposition as an optimization problem,
with correctness guarantees and a clear reliability objective {\bf (Section~\ref{sec:problem})}.
This precise formulation allows the optimization space, objective, and search to be improved independently,
beyond the solutions that this paper presents.

\item  {\bf A model of the validation reliability:} 
We model how reliably a validator judges a candidate query
through decomposed tasks from their SQL
complexity, interface load, as well as coherence and relevance to the natural-language question.
The overall optimization objective balances the reliability of each task against the number of judgments.
We present a first instance of this model targeting human validators {\bf (Section~\ref{sec:hardness})};
alternatives can be readily plugged into our optimization framework.

\item {\bf Search space and strategy:}
We show how a rich space of decompositions suitable for validation can be
efficiently enumerated by applying sound rewrite rules to the candidate query's logical plan,
and by picking appropriate ``cut points'' to partition it,
such that every decomposition provably recomposes to the candidate query {\bf (Section~\ref{sec:space})}.
We present an exact search algorithm that finds the optimum set of decomposed validation tasks
by dynamic programming or, when tasks can be potentially shared, by A* {\bf (Section~\ref{sec:search})}.

\item {\bf A user study with human validators:}
We conducted a user study with 20 SQL-proficient students (volunteers among 218 we invited) in a supervised environment where only their validation skills (and no LLMs/Internet) were used. Each of them validated one LLM-generated
candidate as a whole query and another decomposed by \sys{}. Decomposition increased the share of errors caught
from 36\% to 91\%, with no
false rejection, and they located the error in 9 of 10 rejections
{\bf (Section~\ref{sec:userstudy})}.

\item {\bf Experiments with LLM validators:}
Although our primary motivation is to help human validators, it also helps LLM validators. On 305 questions
from the BEAVER~\cite{chen2026beaver} and Spider~2.0~\cite{lei2024spider}
benchmarks,
\sys{} 
raises the error-detection recall of
three LLMs by 7.0--16.9 percentage points compared with
whole-query validation, while keeping precision above 90\%. 
Therefore, any text-to-SQL method can benefit from \sys{} as an additional checking mechanism
{\bf (Section~\ref{sec:evaluation})}.
\end{enumerate}

\section{Related Work}
\label{sec:related-work}

\emph{Semantic correctness of SQL.}
Alrashed et al.\ benchmark three algorithms for reference-free validation:
SQL critique (direct LLM judgment), round-trip critique (back-translation into
a question), and synthetic execution (LLM-generated test
cases)~\cite{alrashed2026verification}. Other
methods detect semantic errors with database probes and LLM
signals~\cite{gong2025sqlens,tian2026pvsql} or with trained
models~\cite{chen2023errordetection,chen2025sqlcritic,qiu2025herosql}, or
abstain when a query is
uncertain~\cite{chen2025rts,chen2026rtsplusplus}; NL2SQL-Bugs benchmarks such
errors~\cite{liu2025nl2sqlbugs}. LLM judges~\cite{zheng2023judging} comparing two queries
often miss a changed condition~\cite{rahaman2025sqlunderstanding}. These techniques improve
the evidence for one judgment and are orthogonal to \sys{}, which structures
the judgment itself and can apply each of them to a block
(Section~\ref{subsec:exp-effectiveness}).

\emph{Validation inside text-to-SQL generation.}
Text-to-SQL systems also validate implicitly, to choose or repair a
query: they drop queries that fail or return no
rows~\cite{wang2018executionguided}, vote over or cross-check execution
results~\cite{wang2023selfconsistencyimproveschainthought,dong2023c3,li2025alphasql,yang2025sqldriller},
rank candidates with trained selectors or reward
models~\cite{ni2023lever,pourreza2024chasesql,liu2026xiyansql,zhang2026rewardsql},
or let the LLM correct itself~\cite{pourreza2023dinsql,shinn2023reflexion}.
Such pipelines usually return their best candidate even when every candidate
is wrong~\cite{alrashed2026verification}.
\sys{} judges a single candidate without
alternatives or execution, so a pipeline can use it to accept or reject the
query it returns.

\emph{Decomposition for SQL generation.}
DIN-SQL and DTS-SQL decompose the generation of SQL into stages such as schema
linking~\cite{pourreza2023dinsql,pourreza2024dtssqldecomposedtexttosqlsmall},
QPL into the steps of a query plan~\cite{eyal2023qpl}, and CHASE-SQL into
sub-questions~\cite{pourreza2024chasesql}; these methods decompose to
\emph{produce} a query. Outside SQL, verifying the steps of mathematical
reasoning is more reliable than verifying final answers or whole
chains~\cite{lightman2024verify,ling2023deductive}. We decompose a fixed
candidate to \emph{validate} it: the decomposition must preserve the candidate
exactly, errors included, and its granularity is chosen for a validator rather than for a generator.

\emph{Helping people read and validate queries.}
Explaining a query in natural language lets users check that it
matches their intent~\cite{simitsis2009talkback,koutrika2010explaining}, an
idea that round-trip methods automate with
LLMs~\cite{allamanis2024roundtrip,alrashed2026verification}. Interactive interfaces explain queries
step by step, ask users about uncertain parts, or show sample
data~\cite{elgohary2020splash,tian2023,tian2024sqlucid,yao2019misp,narechania2021};
QueryVis and relational diagrams visualize nested
queries~\cite{leventidis2020queryvis,gatterbauer2024}; and reference-based
tools explain wrong queries with counterexamples, abstract instances, repair
hints, and provenance~\cite{miao2019ratest,gilad2022cinsgen,hu2024qrhint,hu2026irex}.
Scrutinizer asks claim checkers about one query fragment at a time and plans
its questions with a cost model of their effort~\cite{karagiannis2020scrutinizer}.
Step-by-step explanations follow the candidate's clauses as written, so they
cannot control what each step looks like. \sys{} chooses the blocks the
validator prefers from equivalent forms of the candidate and gives each its
own requirement, so the block verdicts decide the query.

\emph{Equivalence checking and testing.}
When a reference query exists, QED and SQLSolver prove equivalence for
supported fragments~\cite{wang2024qed,ding2023sqlsolver}; VeriEQL,
SpotIt, and SpotIt+ search bounded instances, including under integrity
constraints~\cite{he2024verieql,klopfenstein2026spotit,tremante2026spotitplus};
ParSEval and distilled test suites generate distinguishing
databases~\cite{chen2025parsqval,zhong2020semanticevaluationtexttosqldistilled},
as does test-data generation for complex
queries~\cite{somwase2025testdata};
and optimizer-based validation certifies some LLM rewrites and returns Unknown
otherwise~\cite{narasayya2026leveraging}. Equivalence is undecidable in
general~\cite{trakhtenbrot1950}. Benchmark references are themselves
often wrong~\cite{jin2026pervasive}. We need equivalence only between a candidate and
its own recomposition, which holds by construction and is confirmed by checks on the printed SQL (Section~\ref{sec:implementation});
the comparison with intent is left to the validator, at a granularity it can handle.

\emph{Query rewriting and optimization.}
Rule-based rewriting~\cite{pirahesh1992extensible}, memo-based optimizers in the
Volcano and Cascades tradition~\cite{graefe1993volcano,graefe1995cascades}, and
join enumeration without cross products~\cite{moerkotte2006analysis,moerkotte2008dynamic}
search equivalent plans for execution cost; equality
saturation~\cite{tate2009equality,willsey2021egg,wang2020spores,zhang2023egglog} represents such
spaces compactly. We reuse the memo but optimize a
different objective: how reliably the blocks are judged, with a repeated task charged
once. This resembles multi-query optimization, where a shared subexpression is
computed once~\cite{sellis1988multiple,roy2000efficient} and optimal plan
selection is NP-hard~\cite{sellis1990multiple}. Our optimizer confines
exponential search to the parts of a query where a task repeats. 

\emph{Text-to-SQL benchmarks and models of difficulty.}
Workload studies measure query complexity by length and distinct
operators~\cite{jain2016sqlshare,vashistha2016}, and Spider and BIRD grade
questions by hardness~\cite{yu2019spider,li2024can}. Spider~2.0 grades its
tasks by the length of the gold query~\cite{lei2024spider}, and BEAVER labels
each query by its structure and by whether it requires domain
knowledge~\cite{chen2026beaver}. None grades a validation task for a
particular validator, as our item-response model does.

\section{The \sys\ Framework}
\label{sec:problem}

The input of \sys\  consists of a natural-language question~$Q$, a database schema~$S$ with constraints~$\Sigma$ (e.g., keys and foreign keys), and a candidate SQL
query~$q$ over~$S$. A \emph{validator}, a person or an LLM, 
judges whether $q$ correctly answers $Q$ according to the validator's intended interpretation. 
Fix the validator's intended interpretation of~$Q$. Let $\mathcal{G}(Q)$ be
the set of queries over~$S$ that return the intended answer on every
database instance satisfying~$\Sigma$. A candidate $q$ is \emph{correct} for
$Q$ if $q \in \mathcal{G}(Q)$.
A \emph{validation task} $T = (Q, S, \Sigma, q)$ asks
for a verdict---accept or reject---on whether $q$ is correct for~$Q$.

Note that the set $\mathcal{G}(Q)$ is never materialized: neither the system nor the
validator has a reference query. 
Correctness validation is thus a semantic judgment, not a mechanical equivalence check.

\subsection{Validation by Decomposition}
\label{subsec:validity}

Decomposition replaces one judgment of~$q$ against~$Q$ with local
judgments, one for each block.
In Figure~\ref{fig:q17}, the
validator of B3 checks that its SQL returns the lineitems of eligible parts
whose quantity is below their part's threshold, taking the eligible parts from
B1 and the thresholds from B2 as given. The validator thus reasons about one
step of the query at a time. The correlated subquery, a construct on which
people often err~\cite{ahadi2015}, becomes block B2, whose parameter and
output are explicit.

{\bf Blocks and decompositions.}
A \emph{block} is a SQL query over base tables and named relations produced by
other blocks (its \emph{inputs}), with optional \emph{parameters} supplied by enclosing queries when the block is a correlated subquery.
A \emph{decomposition}~$D$ of $q$ organizes blocks into a rooted directed acyclic graph (DAG), with edges from
\emph{producer blocks} to \emph{reader blocks} that read their results as inputs.
Each input has one producer, and the graph has a
single sink, called the \emph{root block}, that computes the final output.
$\Compose(D)$ is a query obtained by substituting each producer's query for its
references in dependency order, binding parameters at each call site.
\begin{definition}[Semantics-preserving decomposition]
\label{def:valid}
We write $q \equiv_{\Sigma} q'$ when
the two queries produce the same result on every instance of~$S$
satisfying~$\Sigma$, including row multiplicities and, if specified, output ordering.
A decomposition $D$ of $q$ is \emph{valid} if $\Compose(D) \equiv_{\Sigma} q$.
\end{definition}

{\bf Block-wise validation and composite verdicts.}
Each block $b$ in a valid decomposition $D$ of $q$ defines a separate \emph{validation
task}, $T_b=(Q_b,S_b,\Sigma_b,q_b)$. Here, $q_b$ is the block's SQL, and $Q_b$
is its natural-language requirement. The schema $S_b$ describes its base
tables, input relations, and parameters; $\Sigma_b$ records the constraints
that hold on $S_b$.
\sys\ presents $S_b$ and $\Sigma_b$ to the validator as the context for judging $q_b$ against $Q_b$.
We call the tasks of~$D$ \emph{consistent} if (1)~for every block~$b'$
that reads a block~$b$, the description of~$b$'s output in~$S_{b'}$ matches~$Q_b$;
(2)~for every block~$b$, $\Sigma_b$ follows from~$\Sigma$ and the requirements
of the blocks that $b$ reads, directly or indirectly; and (3)~the requirement
of the root block is~$Q$.

A block $b$ is \emph{locally correct} if, on every
database instance and parameter binding satisfying~$\Sigma_b$, its SQL $q_b$
satisfies its natural-language requirement $Q_b$ \emph{assuming that its inputs are given as described}.
This assumption is important, as it allows judgment of $T_b$ to proceed independently
of validation of $b$'s input blocks.
The validator judges every task and returns accept or reject with a reason. The
\emph{composite verdict} accepts $q$ iff every task is accepted; a rejection
names the block and the requirement it violates.

\begin{proposition}[Compositional soundness]
\label{prop:soundness}
Let $D$ be a valid decomposition of $q$ for $Q$ with consistent validation tasks.
If every block of $D$ is locally correct, then $q$ is correct for~$Q$.
\end{proposition}

Proposition~\ref{prop:soundness} follows directly from the definitions
of validity, consistency, and local correctness. Each local judgment is self-contained: it needs the block's SQL, its
requirement, and the descriptions of its inputs, and no verdict on another
block. By contraposition of
Proposition~\ref{prop:soundness}, if the tasks are consistent,
every incorrect candidate has at least one block that is not locally correct, so a validator
that judges every task correctly rejects every incorrect candidate and identifies a locally incorrect block. 

\subsection{The Decomposition Optimization Problem}
\label{subsec:problem}

\begin{figure}[t]
\centering
\begin{tikzpicture}[
  font=\scriptsize,
  stage/.style={draw, rounded corners=2pt, align=center, minimum height=9mm,
                text width=23mm, inner sep=2pt, fill=white},
  arr/.style={-{Stealth[length=1.5mm]}, semithick}
]
\node[stage] (search) {\textbf{Search}\\ optimal\\ decomposition\\ (Section~\ref{sec:search})};
\node[stage, left=4.5mm of search] (space) {\textbf{Define space}\\ valid decompositions\\ of $q$\\ (Section~\ref{sec:space})};
\node[stage, right=4.5mm of search] (obj) {\textbf{Objective}\\ probability of a\\ correct verdict\\ (Section~\ref{sec:hardness})};
\node[stage, below=5mm of search] (val) {\textbf{Validate}\\ judge blocks,\\ combine verdicts\\ (Section~\ref{subsec:validity})};
\draw[arr] (space) -- (search);
\draw[arr] (obj) -- (search);
\draw[arr] (search) -- (val);
\node[above=2.5mm of space, align=center] (in) {candidate $q$, schema $S$,\\ constraints $\Sigma$};
\draw[arr] (in) -- (space);
\node[above=2.5mm of obj, align=center] (th) {question $Q$,\\ validator profile $\theta$};
\draw[arr] (th) -- (obj);
\node[align=center] (v) at (space |- val) {validator\\ (person or LLM)};
\draw[arr] (v) -- (val);
\node[align=center] (out) at (obj |- val) {verdict,\\ defective block};
\draw[arr] (val) -- (out);
\end{tikzpicture}
\caption{The pipeline of \sys{}. The search
picks the decomposition that maximizes the validator's probability of a
correct verdict, discounted by the number of judgments, and the validator then
judges its blocks.}
\Description{DecoVal's pipeline: the space of valid decompositions and the
objective, the probability of a correct verdict, which takes the question and
the validator profile, both feed the search; the validator judges the blocks
of the chosen decomposition, which yields a verdict and the defective block.}
\label{fig:overview}
\end{figure}
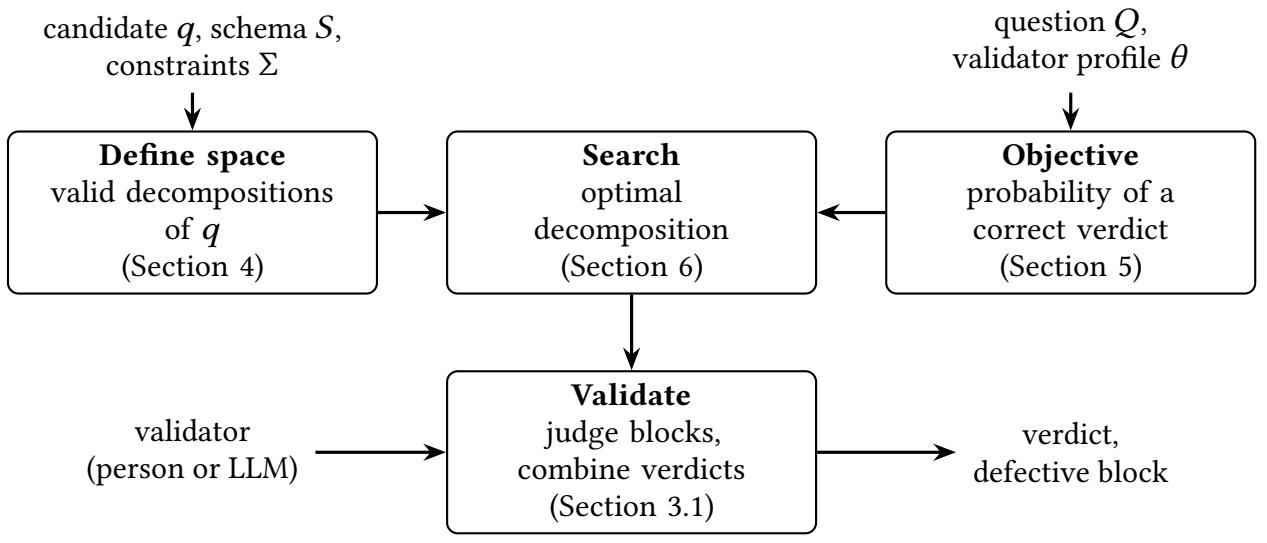

A good decomposition should have four properties. It is \emph{valid} (Definition~\ref{def:valid}),
so that judging by parts enables judging the whole query. Its blocks are
\emph{simple}, each easy for the validator to understand, and \emph{meaningful}, each computing a subgoal of the
question that can be stated as a requirement. It involves \emph{few judgments}:
every block adds a task to judge, and a task that occurs at several places is
judged once.
While validity is a hard constraint, the other properties involve trade-offs.
For example, fewer judgments may lead to more complex blocks, each of which can be harder to validate.
Their balance depends on the validator, whose reliability depends on the complexity of each task.
Ultimately, we are interested in finding a decomposition that maximizes the reliability of the overall judgment.
To model this trade-off and dependency on validators, we
let $\theta$ denote a validator profile and $U_{\theta}(D)$ a modeled measure of validation reliability for
decomposition~$D$. We seek a valid decomposition that maximizes this measure
among all candidates.

\begin{definition}[Validation-aware Query Decomposition]
\label{def:problem}
Given a validation task $T=(Q,S,\Sigma,q)$, a validator profile $\theta$,
and a nonempty finite candidate set $\mathcal{D}(q)$ of valid decompositions of $q$,
find
\begin{equation}
D^{*} \in \arg\max_{D\in\mathcal{D}(q)} U_{\theta}(D).
\end{equation}
\end{definition}

\sys{} solves this problem with the pipeline of Figure~\ref{fig:overview},
which parallels the key components of traditional query optimization (but for a different objective):
plan enumeration, cost estimation, and search procedure.
For \emph{plan enumeration}, we define the space of valid decompositions
by applying sound rewrite rules to the candidate query's logical plan and finding a partitioning of plan operators into blocks.
Instead of \emph{cost estimation}, our objective~$U_{\theta}$ measures how reliably the validator judges the tasks (Section~\ref{sec:hardness}).
We score each task with an item-response model~\cite{rasch1960probabilistic,reckase2009mirt},
in which the chance of a correct judgment falls
when the block's SQL complexity and interface load (the inputs and parameters it reads)
exceed the validator's preferences and capabilities (as specified by $\theta$),
and if the block's natural-language requirement feels disconnected from~$Q$.
As validating the whole query requires judging all blocks in a decomposition,
our model also penalizes the total number of judgments.
As for the \emph{search procedure}, we employ two methods (Section~\ref{sec:search}) for finding optimal decompositions:
dynamic programming for decompositions in which no validation task can be reused,
and A* for the general case (e.g., when the same filtered join can be cut out of two branches of a union).

\emph{Human vs.\ LLM validators.}
We want \sys{} to work for human validators, who we believe are ultimately responsible for interpreting the query intent.
Hence, our modeling of $U_{\theta}$ in Section~\ref{sec:hardness} is largely motivated by prior work on human reliability.
Nonetheless, \sys{} is designed to see a validator only through its $U_{\theta}$ and verdicts,
so LLM validators can be used just as easily.
Our user study in Section~\ref{sec:userstudy} tests how \sys{} helps human users:
whether blocks are easier to judge than whole queries, and how decomposition improves overall effectiveness of human validators.
Section~\ref{sec:evaluation} further evaluates the effect of \sys{} with LLM validators,
which are cheaper and allow larger-scale evaluation involving hundreds of candidate queries.
As we will see in Section~\ref{sec:evaluation}, \sys\ also helps in this setting,
significantly improving other LLM-based validation methods.

\section{The Decomposition Space}
\label{sec:space}

The candidate set $\mathcal{D}(q)$ of Definition~\ref{def:problem}
determines which blocks the validator can receive.
As a query optimizer considers plans beyond the written one, \sys{}
expands the candidate's plan with sound rewrite rules into a data structure of
equivalent subplans (Section~\ref{subsec:plans}) and forms blocks from these
subplans at cut points (Section~\ref{subsec:blocks}). Every decomposition
obtained this way is valid, and the resulting space $\Dspace$ serves
as~$\mathcal{D}(q)$.

\subsection{Plans, Rewrites, and the Memo}
\label{subsec:plans}
\label{subsec:memo}

In contrast to traditional query optimization for execution performance,
where the space of execution plans can include those that do not resemble the original query at all,
we take a more deliberate approach in enumerating decompositions,
because it is beneficial for the validator to recognize the connection between the candidate query and the validation tasks.
Hence, \sys{} considers only the plans that reorder the candidate's inner
joins and move its filters within defined regions, preserving the main query structure.
In particular, \sys{} never decorrelates a subquery;
instead, it treats a block cut from a correlated subquery as parameterized,
i.e., receiving values of its external column references as named input parameters (Section~\ref{subsec:blocks}).

\emph{Skeleton plans.} \sys{} imports the candidate query's logical plan into a
\emph{skeleton plan} defined using the following operators:
$\mathit{Scan}(T,o)$ reads an \emph{occurrence}~$o$ of a base table~$T$;
$\mathit{Join}(L,R,J)$ is an inner join with unordered inputs and a set~$J$ of \emph{conjuncts} (more below);
$\mathit{Filter}(P,c)$ applies one conjunct~$c$;
other unary and binary operators, which we generally refer to as \emph{peg} operators
(which mark regions for rewriting, as described below), including:
aggregation, generalized projections, \sqlkw{DISTINCT},
\sqlkw{ORDER} \sqlkw{BY}, \sqlkw{LIMIT}, windows, outer-, semi-, and anti-joins, set/bag operations,
and references to CTEs (defined by \sqlkw{WITH}) and derived tables (subqueries in \sqlkw{FROM}).
Note that an \emph{occurrence} is one \sqlkw{FROM} item;
multiple references to the same base table or CTE remain distinct occurrences.
For example, in Q17, \texttt{l} and \texttt{li} are two occurrences of \texttt{lineitem}.

Normalization splits conjunctive query predicates into \emph{conjuncts} (while keeping \sqlkw{OR} as atomic).
We mark external column references in correlated subqueries as \emph{free variables}, but otherwise allow them in expressions;
i.e., no decorrelation is done.
We also drop unnecessary projections that only remove or rename columns.
We write $P_0$ for the normalized plan of~$q$.

\emph{Region-based rewriting.}
Rewrites operate within each \emph{region}, a maximal
connected part of~$P_0$ made of joins and filters within one \sqlkw{SELECT} statement.
A region stops at its \emph{inputs}, which are the scans and the subplans rooted at peg operators directly below it.
The body of a subquery (which creates a new SQL scope) in a predicate or \sqlkw{SELECT} list also starts a region of its own.
Rewrites do not move joins or filters moves out of their current region.
For example, an aggregation (which is a peg operator) would separate the joins and filters below it from any filter above.

Within each region, \sys{} uses the join-reordering and filter-placement rules of query
optimizers~\cite{selinger1979access,pirahesh1992extensible}: a region's joins
can be reordered in any way that creates no cross product, and each local
conjunct can be applied at any point above its owner.
The set $\mathcal{R}$ of rewrite rules is`  ` as follows:
\begin{itemize}[leftmargin=*]
\item[\textbf{R1}] $A \bowtie B \equiv B \bowtie A$ (join inputs are unordered);
\item[\textbf{R2}] $(A \bowtie B) \bowtie C \equiv A \bowtie (B \bowtie C)$, applied only if
the new inner join is spanned by some conjunct, so that no cross product is
created;
\item[\textbf{R3}] $\mathit{Filter}_{c_1}(\mathit{Filter}_{c_2}(A)) \equiv
\mathit{Filter}_{c_2}(\mathit{Filter}_{c_1}(A))$;
\item[\textbf{R4}] $\mathit{Filter}_{c}(A) \bowtie B \equiv \mathit{Filter}_{c}(A \bowtie B)$
if $c$ (sans free variables) can be evaluated over $A$ alone.
\end{itemize}
If every conjunct of~$q$ is deterministic and raises no runtime
error, each rule preserves the output bag on every instance and every
binding of the free variables.

Finally, within each region, \sys{} ensures that a join conjunct is pushed down as much as possible:
i.e., it is applied at the \emph{lowest} join whose subtree contains the inputs it needs (sans free variables).
On the other hand, unlike traditional query optimization, \sys{} does not always push non-join filters down:
intuitively, validating a join first and then a filter is not necessarily harder than
validating a filtered table first and then a join involving it.

In general, rewrites expose subplans that the skeleton plan lacks without changing the shape of the candidate's SQL much;
each exposed subplan can potentially become a block.
By exploring different rewrites, \sys{} can, for example,
break down a complex join involving many tables into simpler joins,
or find opportunities for sharing the effort of validating common subplans.

\emph{The memo.}  
\sys{} compactly represents the space of all rewritten plans in a \emph{memo}~\cite{graefe1995cascades} of \emph{groups}: two
subplans share a group exactly when R1--R4 rewrite one into the other.
A group holds one or more subplans that accomplish the same computation but in different ways.
Since \sys{} rules only reorder joins and move filters within a region,
all subplans of a group read the same set of occurrences, apply the same conjuncts,
and apply the same peg operators to the same child groups, but using different query expressions.
For a group~$g$, let $\mathit{plans}(g)$ denote the subplans obtained by choosing a query expression for~$g$
over its chosen recursively child subplans.
For the root group~$g_0$, which contains~$P_0$, these are exactly the plans reachable from~$P_0$ by R1--R4 across all regions.
The memo is \emph{sound}: all subplans in a group are logically equivalent.

\subsection{Decomposition by Cut Points}
\label{subsec:blocks}

Given a plan represented by the memo, we can further enumerate its valid decompositions (Section~\ref{subsec:validity})
by enumerating a set of \emph{cut points} selected from the memo groups representing its subplans.
Each selected cut point marks a block output, which becomes a named input to its readers;
the plan fragments between these boundaries form the blocks.
Hence, we can identify each block by a pair $(g,I)$, where $g$ is the memo group corresponding to the block's output,
and $I$ is the set of memo groups corresponding to its inputs.

Any subplan can be a cut point, subject to three rules.
(1)~Every block contains an operator other than scans, references, and column
selection. The motivation for this rule is that every task should judge some computation;
a bare scan or reference is too trivial to be a block by itself.
(2)~Within one \sqlkw{SELECT} statement, a cut may separate the
\sqlkw{FROM}/\sqlkw{WHERE} part from the operators that follow it, as the
boundary between B3 and B4 in Figure~\ref{fig:q17} does, but never falls among
operators following \sqlkw{FROM}/\sqlkw{WHERE}---aggregation, \sqlkw{HAVING}, select list,
\sqlkw{DISTINCT}, \sqlkw{ORDER} \sqlkw{BY}, and \sqlkw{LIMIT}.
The motivation is that from our experience,
cutting into post-\sqlkw{FROM}/\sqlkw{WHERE} steps is either unnecessary
(e.g., substeps such as \sqlkw{DISTINCT} and \sqlkw{ORDER} \sqlkw{BY} are too small to warrant separate tasks)
or too difficult (e.g., the result of grouping, separated from aggregation, cannot even be described as a flat table for verification).
On the other hand, it is beneficial to consider cutting a complex \sqlkw{FROM}/\sqlkw{WHERE} part into simpler blocks.
(3)~The body of a CTE referenced multiple times in the candidate query is cut in every decomposition whenever it can be a block.
Such a \emph{forced cut} keeps the boundary that the candidate query already clearly draws;
hence, the validator only needs to validate this CTE once as a separate subproblem.

\emph{Parameterized blocks.}
\sys{} makes it possible to create a block whose subplan contains free variables.
This flexibility allows \sys{} to validate subqueries independently from the containing query.
In this case, the free variables become explicit input parameters of the block;
\sys{} will describe the meaning of these input parameters to the validator,
enabling them to validate the block as a parameterized query instead of a standalone one. 
For example, in Figure~\ref{fig:q17}, block B2 holds the threshold subquery
with its free variable \texttt{p.p\_partkey} replaced by a parameter~$x$, so $\mathrm{B2}(x)$ returns the threshold of
part~$x$; B3 calls \texttt{B2(B1.p\_\_p\_partkey)} for each of its rows, binding $x$ to that row's part key.
In both contexts, B2 is readily interpreted as a function.

It is also worth noting that subqueries (correlated or not) do not automatically imply a new block or validation task.
Indeed, more capable validators may tolerate or even prefer blocks containing some subqueries.
The choice is up to \sys{}'s decomposition objective (Section~\ref{sec:hardness}).

\emph{Search space.}
\label{def:space}
Let $\Dspace$ be the set of decompositions that the rules~$\mathcal{R}$ and the
cut points yield.
Each decomposition has one root block at $g_0$ and one producer per selected
boundary, with all blocks reachable from the root.

\begin{proposition}[Validity of the cut construction]
\label{thm:validity}
If every conjunct of~$q$ is deterministic and raises no runtime error, then for every decomposition
$D \in \Dspace$, $\Compose(D)$ and~$P_0$ return the same result on every
instance, so $D$ is valid in the sense of Definition~\ref{def:valid}.
\end{proposition}

\section{Modeling
Decomposition Objective}
\label{sec:hardness}

The objective $U_\theta$ plays the role of a query optimizer's cost
model: it rates every decomposition in~$\Dspace$ by how reliably a validator
is expected to judge it. Section~\ref{subsec:sources} first describes
how various features of a block can affect validator reliability.
Section~\ref{subsec:reliability} then shows how to determine validator reliability
for a given block using its features and the validator's profile.
Section~\ref{subsec:objective} discusses how to combine reliability across blocks
into an overall optimization objective.

\subsection{Block Features}
\label{subsec:sources}
\label{subsec:features}

A block can be hard to validate for various reasons;
we break them down into three categories in Table~\ref{tab:features},
further discussed below.

\emph{SQL complexity.}
Every construct inside a block is a decision the validator must check against the
intent: which rows a join pairs, which rows a predicate keeps, what a group
contains, whether duplicates and \sqlkw{NULL}s are handled as intended, and which
outer row a correlated subquery refers to.
The features count these constructs, one for each kind on which people and LLMs are known to err.
Join omissions and wrong join conditions are among the most frequent
logical errors~\cite{taipalus2018,taipalus2019}; correlated subqueries and
self-joins have the lowest success rates~\cite{ahadi2015}; negation and
\sqlkw{NULL} semantics mislead even database professionals~\cite{toussaint2022};
and mixing Boolean connectives causes nesting errors~\cite{greene1990,taipalus2018}.
LLMs err most on set operations, subqueries, and joins~\cite{ning2024,li2024dawn}.
Finally, query length and operator diversity serve as proxies for the overall
load~\cite{jain2016sqlshare,peitek2021}.

\emph{Interface load.}
A block $b$ replaces part of the query with named input relations and parameters.
The validator no longer sees how an input relation is computed or where in an outer block a parameter comes from;
instead, the validator must now understand the meaning of such input relations and parameters in $S_b$,
as part of the validation task $T_b=(Q_b,S_b,\Sigma_b,q_b)$.
Furthermore, the validator must consider additional constraints in $\Sigma_b$ associated with these input relations and parameters,
which \sys{} propagates across block boundaries (Section~\ref{sec:implementation}).
Therefore, we include the counts of input relations and parameters as
measures of load placed on the validator to understand these interfaces.

\emph{Semantic score.}
Structural measures cannot tell whether a block computes a meaningful intermediate concept.
For example, in Q17, applying the brand and container filters to
\sqlkw{part} or to an input of already-joined \sqlkw{part}-\sqlkw{lineitem} pairs gives two blocks with the same two filtering predicates.
The former directly produces the eligible-parts concept named by the question.
A validator judges a block by mapping it to the intent,
so it needs blocks whose outputs have an interpretable meaning (\textsf{coherence})
and are relevant to the overall question (\textsf{relevance}).
Given a block $b$, \sys\ uses an LLM to obtain these two measures from its SQL $q_b$.
\textsf{Coherence}, rated without the overall question, asks whether
the relation makes sense on its own: 3 for a natural unit with a clear name
and a clear row, such as the eligible parts; 2 if the name needs a
qualification or some columns serve only a later step; 1 if the relation is
describable only as an operation, such as the rows of one table joined with
another, or its row is unclear; and 0 if it makes no sense alone, such as half of a condition.
\textsf{Relevance} rates the relation against the overall question:
\emph{complete} if it computes a concept that the question names, with every condition the
question attaches to it; \emph{narrower} if some of these conditions are still
to come; \emph{technical} for a preparatory step that matches no phrase of the
question; \emph{split} if it applies a condition of the question only in part,
such as one end of a range; and \emph{unrelated} if it has no connection to
the question. On each scale, the top level has factor~$1$, the second level, which still marks a sound unit,
has factor $1-\varepsilon$ with $\varepsilon=0.05$, and the $k$-th level for
$k\ge3$ has factor $2^{2-k}$. The semantic score $s(b)\in(0,1]$ of a block $b$ is the
product of the coherence and relevance factors
(the root block, whose natural-language requirement is the overall question, always has a semantic score of $1$).

\begin{table}[t]
\centering
\footnotesize
\caption{Block features.}
\label{tab:features}
\begin{tabularx}{\columnwidth}{@{}l X@{}}
\toprule
\multicolumn{2}{@{}l}{\emph{SQL complexity}} \\
\textsf{inner\_join} & inner joins ($m-1$ for an $m$-input region) \\
\textsf{outer\_join} & left, right, and full outer joins \\
\textsf{pred} & filter, join, \sqlkw{HAVING}, and \sqlkw{ON} conjuncts \\
\textsf{bool\_mix} & \sqlkw{OR} and \sqlkw{NOT} operators inside conjuncts \\
\textsf{negation} & \sqlkw{NOT} \sqlkw{IN}, \sqlkw{NOT} \sqlkw{EXISTS}, \sqlkw{<> ALL}, negated predicates \\
\textsf{null\_sensitive} & \sqlkw{IS} [\sqlkw{NOT}] \sqlkw{NULL}, \sqlkw{COALESCE}, \sqlkw{NULLIF}, \sqlkw{NOT} \sqlkw{IN}, outer-join columns \\
\textsf{distinct} & \sqlkw{DISTINCT}, \sqlkw{DISTINCT} \sqlkw{ON}, aggregates over \sqlkw{DISTINCT} \\
\textsf{agg} & grouping and global aggregation operators \\
\textsf{expr} & function calls, \sqlkw{CASE} branches, arithmetic, casts \\
\textsf{setop} & \sqlkw{UNION}, \sqlkw{INTERSECT}, \sqlkw{EXCEPT} operators \\
\textsf{order\_limit} & \sqlkw{ORDER} \sqlkw{BY} and \sqlkw{LIMIT}/\sqlkw{OFFSET} operators \\
\textsf{table\_reuse} & repeated uses of one relation (self-joins) \\
\textsf{uncorr}, \textsf{corr} & uncorrelated and correlated subquery calls \\
\textsf{depth} & subquery nesting depth inside the block \\
\textsf{n\_optypes} & distinct operator types \\
\textsf{length} & tokens of the block \\
\midrule
\multicolumn{2}{@{}l}{\emph{Interface load}} \\
\textsf{n\_relations} & distinct base tables and block inputs read \\
\textsf{param} & outer columns the block depends on \\
\midrule
\multicolumn{2}{@{}l}{\emph{Semantic score} (combined from below into $s(b) \in (0,1]$)} \\
\textsf{coherence} & whether the block's output is meaningful on its own  \\
\textsf{relevance} & whether the block's goal relates to the overall question \\
\bottomrule
\end{tabularx}
\end{table}

\subsection{Validator Profile and Reliability Model}
\label{subsec:reliability}

The validator profile~$\theta$ captures the validator's preferences and/or capabilities:
how much SQL of each kind, such as joins, predicates, or nested subqueries, it wants to judge within one block.
Let $X_\theta$ denote the set of block features in Table~\ref{tab:features} except \textsf{coherence} and \textsf{relevance};
$\theta$ assigns a threshold $\theta(x)\ge0$ to every feature $x \in X_\theta$.
For example, one validator may be able to read a block joining four tables with ease,
while another wants at most two-way joins;
their profiles would assign different thresholds for \textsf{inner\_join}.
Section~\ref{sec:introduction} gives a similar example for the correlated threshold of Q17.
Generally speaking, raising a threshold admits larger and/or more complex blocks.
The best profile will differ across validators, human and LLM alike:
our human study (Section~\ref{sec:userstudy}) reveals that some participants preferred smaller blocks while skilled ones preferred larger ones;
our experiments with LLM validators (Section~\ref{subsec:exp-granularity}) similarly show that
the best block granularity depends on the LLM.

Let $v_x(b)$ denote the value of feature $x \in X_\theta$ for block $b$, and let $s(b)\in(0,1]$ denote its semantic score.
Given a validator profile~$\theta$, the normalized excess of $b$ in feature $x$ is:
\[
\mathit{excess}_x(\theta, b)=
\frac{\max\{0,\,v_x(b)-\theta(x)\}}{\max\{\theta(x),1\}}.
\]
For constants $\beta>0$ and $\gamma\ge0$ of
the model, the reliability of~$b$ is
\begin{equation}
\label{eq:reliability}
F(\theta,b)=
\underbrace{s(b)^\gamma}_{\text{Semantic factor}}\,
\underbrace{\prod_{x\in X_\theta}
\exp\bigl(-\beta\,\mathit{excess}_x(\theta, b)\bigr)}_{e^{-R(\theta,b)}\text{ for structural risk }R(\theta,b)}.
\end{equation}
Here $R(\theta,b)=\beta\sum_{x\in X_\theta}\mathit{excess}_x(\theta,b)$ is the \emph{structural risk} of~$b$.
For each feature $x \in X_\theta$, beyond the validator's threshold $\theta(x)$,
an excess in feature value lowers validator reliability at a rate set by~$\beta$;
normalized excess lets one decay serve features of different scales.
If every feature value is within the profile, $R(\theta,b) = 0$.
Note that $R(\theta,b)$ is a conjunctive, non-compensatory multidimensional item-response model~\cite{reckase2009mirt},
where slack on one dimension does not offset an excess on another:
e.g., misreading a single join or \sqlkw{NULL} test suffices to misjudge a block.
Finally, the exponent $\gamma$ weights a block's semantic score against its structural risk.

\subsection{Optimization Objective}
\label{subsec:objective}
\label{subsec:distinct}
\label{subsec:classes}

A block that is used multiple times as input in a decomposition only needs to be judged once.
More generally, \sys{} detects when two blocks can be associated with the same task:
if they are equivalent up to a renaming of aliases, the verdict on one transfers to the other.
We call the distinct tasks of a decomposition~$D$ its \emph{task classes}~$\classes(D)$.
Since all blocks of a class~$c$ are essentially equivalent,
we will with a slight abuse of notation use $c$ in place of $b$ in $F(\theta,c)$, $R(\theta,c)$, $s(c)$, etc.

The reliability of a decomposition multiplies the reliabilities of its distinct
tasks and discounts the number of judgments it demands:
\begin{equation}
\label{eq:objective}
U_{\theta}(D) \;=\; e^{-\mu|\classes(D)|}
\prod_{c \in \classes(D)} F(\theta, c),
\end{equation}
where $\mu\ge0$ prices one judgment. Maximizing
this objective is equivalent to minimizing the total cost of the task classes:
\begin{multline}
\label{eq:cost}
J_{\theta}(D) \;=\; \sum_{c \in \classes(D)} w_\theta(c),\\
\text{where}\; w_\theta(c) = \mu-\log F(\theta, c) = \mu + R(\theta, c) - \gamma\log s(c).
\end{multline}
The cost $w_\theta(c)$ adds the overhead of one judgment, the structural risk of its blocks, and their semantic cost.

\begin{proposition}[Objective as a reliability bound]
\label{prop:bound}
If the validator's judgments of distinct tasks fail independently,
and each task $c$ is judged correctly with probability $F(\theta, c)$,
then the composite verdict using consistent validation tasks for a valid decomposition is correct
with probability at least $\prod_{c \in \classes(D)} F(\theta, c)$.
\end{proposition}

Intuitively, this objective rewards decompositions that make
each distinct task simpler and therefore more reliable to judge,
it also prices the effort of each additional task.
Consider, for example, a split of $b$ into $b_1,b_2$ that removes $b$'s task class and
introduces two new, distinct classes, with all other class costs unchanged.
The split pays off exactly when
\[
\log F(\theta,b_1)+\log F(\theta,b_2)-\log F(\theta,b)>\mu.
\]
One of the two blocks reads the other and keeps the output of~$b$, and with it
the semantic score of~$b$.
The split therefore pays off only if it lowers the
structural risk by more than~$\mu$ plus the semantic cost of the new block.
Once every block is within the validator's thresholds, no split pays off, and
among splits with equal structural gains, the objective prefers the one that
cuts out the more meaningful block.
A repeated task is paid once, so cutting it out of several places at once can pay off even
when cutting it out of one place does not.

\section{Optimal Decomposition Search}
\label{sec:search}
\label{subsec:nonadditive}

\sys{} finds a decomposition $D^{*}\in\Dspace$ of minimum $J_\theta$ as
a query optimizer finds a plan: it chooses a block for the root group~$g_0$
and, recursively, for every input of a chosen block. If every task class held
a single block, $J_\theta$ (Equation~\eqref{eq:cost}) would add up the costs of
the chosen blocks, and dynamic programming over memo groups would find the
optimum (Section~\ref{subsec:dp}). A class may contain several blocks sharing
the same validation task, however: e.g., when two branches of
a union filter the same join. In this case, the cost of choosing the block on one branch then depends on whether the
other does, similar to shared subexpressions in multi-query
optimization~\cite{sellis1988multiple}. Dynamic programming, which picks the
best decomposition of each group on its own, can miss the optimum. We call
such a class \emph{shareable} and write $\classes_{\mathit{sh}}$ for these
classes; when $\classes_{\mathit{sh}}\ne\emptyset$, \sys{} uses A* search
(Section~\ref{subsec:astar}).

\subsection{DP for No Shareable Classes}
\label{subsec:dp}
When ${\classes_{\mathit{sh}}}=\emptyset$, every block of a decomposition
is in its own class, so $J_\theta(D)$ is the sum of the costs of $D$'s blocks.
Recall from Section~\ref{subsec:blocks} that a block is represented by $(g,I)$,
where $g$ is the memo group representing its output and $I$ the set of memo groups representing its inputs.
Given a target subquery $g$, to find $\mathit{best}(g)$, the minimum cost of decomposing $g$,
we consider $\mathcal{B}$, the set of all candidate blocks in the valid decompositions of~$\Dspace$.
The optimization boils down to picking the best input groups for the block outputting $g$:
\begin{equation}\label{eq:dp}
\mathit{best}(g)=\min_{(g,I)\in\mathcal{B}}
\Big[w_\theta(g,I)+\sum_{i\in I}\mathit{best}(i)\Big].
\end{equation}
Dynamic programming keeps one value $\mathit{best}(g)$ per memo group
(along with the input groups $I_\mathit{best}$ achieving this value), as
in memoized plan enumeration~\cite{selinger1979access,graefe1995cascades}.
We call this procedure \textsc{DP}.

As a further optimization to the DP implementation,
recall from Section~\ref{subsec:blocks} that we force-cut at groups corresponding to shared subqueries.
Such groups can be resolved once and then excluded from DP enumeration and cost tallying in Equation~\eqref{eq:dp};
their total cost can be included in $\mathit{best}(g_0)$ at the end of DP.

\begin{theorem}[Exactness without shareable classes]
\label{thm:dp}
If no class is shareable, DP returns a decomposition in~$\Dspace$ of minimum~$J_\theta$ (proof in Appendix~\ref{app:proof-dp}).
\end{theorem}
\subsection{General A* Search}
\label{subsec:astar}

Algorithm~\ref{alg:search} presents our A*-based algorithm~\cite{hart1968formal} for the general case when some classes may be sharable.
It builds a decomposition one block at a time, from the root down, and
remembers which shareable classes it has already paid for. A search state
$\sigma=(\Pi,\Phi)$ holds the pending groups~$\Pi$, which still need to be further decomposed,
and the shareable classes~$\Phi$ paid so far. The search starts with the root
group and the forced cuts pending and ends when no group is pending. A move
takes the first pending group~$p$, in a fixed topological order, and gives it a
block $(p,I)$ of some class~$c$: $p$ is then replaced by the input groups $I$, $c$ is
added to~$\Phi$ if it is shareable, and the move costs $w_\theta(c)$, or nothing if
$c\in\Phi$. States with the same $(\Pi,\Phi)$ have the same completions, so the
search keeps only the cheapest path to each. A* expands states in order of
$J(\sigma)+h(\sigma)$. Here $J(\sigma)$ is the part of~$J_\theta$ paid so far,
and $h(\sigma)$ estimates the rest: it adds up $\mathit{best}(p)$ over the
pending groups~$p$, computed as if every shareable class were free.

Besides the optimization involving forced cuts used also in DP (Section~\ref{subsec:dp}),
Algorithm~\ref{alg:search} applies two reductions in \textsc{Fold} to every new state to keep the search small.
(1) A pending group below which no unpaid shareable class occurs is solved at once by \textsc{DP} with the classes of~$\Phi$ free, since no
choice inside it affects another group.
(2) A paid class that no pending group can
use is dropped from~$\Phi$, which merges states that differ only in past
choices.

\begin{algorithm}[t]
\caption{Exact decomposition search}
\label{alg:search}
\small
\algtext*{EndWhile}\algtext*{EndFor}\algtext*{EndFunction}
\begin{algorithmic}[1]
\Require blocks $\mathcal{B}$, class costs $w$
\Ensure a decomposition in $\Dspace$ of minimum $J_\theta$
\State \textbf{if} no class is shareable \textbf{then return} the result of \textsc{DP}
\State $Q\gets\{$\Call{Fold}{start state}$\}$
\While{$Q\ne\emptyset$}
  \State take from $Q$ the state $\sigma=(\Pi,\Phi)$ with least $J(\sigma)+h(\sigma)$
  \State \textbf{if} $\Pi=\emptyset$ \textbf{then return} the decomposition of $\sigma$
  \State $p\gets$ the first pending group of $\sigma$
  \For{each block $(p,I)\in\mathcal{B}$, of class $c$}
    \State $\sigma'\gets\sigma$ with $p$ replaced by $I$ (excluding forced cuts) and $c$ paid
    \State add \Call{Fold}{$\sigma'$} to $Q$
  \EndFor
\EndWhile
\Function{Fold}{$\sigma$}
  \For{each $p\in\Pi$ with no unpaid shareable class below it}
    \State solve $p$ by \textsc{DP} with the classes of $\Phi$ free
  \EndFor
  \State drop from $\Phi$ the classes no pending group can use; \Return $\sigma$
\EndFunction
\end{algorithmic}
\end{algorithm}

\begin{theorem}[Exactness of the search]
\label{thm:astar}
The heuristic~$h$ is admissible and consistent, and
Algorithm~\ref{alg:search} returns a decomposition in~$\Dspace$ of
minimum~$J_\theta$ (proof in Appendix~\ref{app:proof-astar}).
\end{theorem}

\begin{proposition}[Cost]
\label{prop:cost}
Given~$\mathcal{B}$, the search evaluates~$F$ once per class, at most
$|\mathcal{B}|$ times. Without shareable classes, \textsc{DP} takes
$O\big(\sum_{(g,I)\in\mathcal{B}}(1+|I|)\big)$ time. With shareable classes,
A* expands at most $2^{|\classes_{\mathit{sh}}|}$ states for each set of
pending groups that remain after folding.
\end{proposition}

The search is thus exponential only in the parts of a query where a task
repeats. Enumerating~$\mathcal{B}$ is the remaining cost: a block pairs the group it computes with a set of input cut points, so $|\mathcal{B}|$ can grow
exponentially with the number of cut points of a region.

\section{Implementation}
\label{sec:implementation}

\sys{} is implemented in Python; 
Figure~\ref{fig:overview} shows its pipeline.
Semantic scoring for the objective and
task assignment for validation call an LLM, the validator judges the resulting
tasks, and the rest of the pipeline is deterministic. All stages before
validation share a per-query time budget; a query that exceeds it receives no
decomposition. Appendix~\ref{app:prompts} lists the prompts.

\emph{Import and memo.}
The importer parses a candidate with SQLGlot~\cite{sqlglot} in the PostgreSQL
dialect into the skeleton plan of Section~\ref{subsec:plans}; it refuses
syntactically invalid queries and queries with constructs that the skeleton plan does not represent, such as
\sqlkw{LATERAL} joins, recursive CTEs, and table functions.
A side table maps every conjunct and operator to its original SQL text.
\sys{}
builds the memo by equality saturation~\cite{tate2009equality} in an egglog
e-graph~\cite{willsey2021egg,zhang2023egglog}; once no rule adds a new plan,
its e-classes are the groups of Section~\ref{subsec:memo}.

\emph{Task classes.}
Two blocks belong to the same task class if, after their aliases are renamed to \texttt{t0}, \texttt{t1}, \ldots{} in a fixed order, they read the same tables and inputs and apply the same conjuncts and operators.

\emph{Scoring.}
For the structural risk, \sys{} counts
the features of Table~\ref{tab:features} directly on a class's plan fragment,
without rendering its SQL. 
The LLM scorer sees each cut point as the SQL~$q_b$ of a block that outputs it, with the block's inputs expanded to base tables, and rates
all of them in two calls, one per scale.

\emph{Rendering SQL for blocks.}
The printer builds each block's SQL from the side-table text, so the
validator reads the candidate's own expressions.
Proposition~\ref{thm:validity} holds for plans in the memo, and three checks
confirm that the printed SQL preserves it. C1 re-imports the SQL printed for
a memo group and requires the result to fall in the same group. C2 inlines
the blocks of a decomposition into their readers, re-imports the result, and
requires it to fall in the root group. C3 requires PostgreSQL to accept
every printed query in a read-only \sqlkw{EXPLAIN} whenever it accepts the
candidate. We say that \sys{} \emph{supports} a query if it imports
the query, returns a decomposition, and the printed SQL passes C1--C3.

\emph{Constraint propagation.}
\sys{} propagates declared schema constraints through memo groups and
derives additional keys under bag semantics. Each operator has a rule that
computes the constraints of its output from those of its inputs. For example,
grouping by plain columns makes those columns a key. In Figure~\ref{fig:q17},
\texttt{p\_\_p\_partkey} is a key of B1, because it is the primary key of \texttt{part}
and a filter keeps the keys of its input. This key keeps B1's join with
\texttt{lineitem} in B3 from duplicating a lineitem: the join condition equates it with
\texttt{l.l\_partkey}, so each lineitem matches at most one row of B1. B2
computes a global aggregate, so it returns at most one row per call, which
makes its use as a scalar subquery in B3 valid. The task of B3 lists both
facts in~$\Sigma_3$. The rules are sound but not complete: every constraint
they derive holds on every database that satisfies the declared constraints,
so a task never states a false guarantee, but they can miss constraints that
hold. Extending the rules to derive more constraints is future work.

\emph{Task assignment.}
For each non-root block of the selected decomposition, an LLM \emph{writer} generates a
requirement~$Q_b$ from~$Q$, the relevant schema, the block's output columns
and parameters, and the requirements and SQL of all its readers; the block's
own SQL and derived keys are withheld. A separate LLM \emph{describer} generates a
description of each non-root block's output relation from a single SQL
statement that composes the block with all its upstream producers; it describes the SQL as written. 
Blocks of one task class share requirement and description, so together with the two scoring calls, a decomposition with $k$ task classes needs $2 + 2(k-1) = 2k$ LLM calls to give the validator $k$ tasks,
however many cut points the query has.

\section{User Study with Human Validators}
\label{sec:userstudy}

\begin{table}[t]
\centering
\footnotesize
\caption{User study results. Time: mean/median. Located: the share of validations of an incorrect candidate that
reject it at its defective block or at a block that reads its
result.}
\label{tab:userstudy}
\setlength{\tabcolsep}{3pt}
\edef\abovesep{\the\aboverulesep}\edef\belowsep{\the\belowrulesep}
\setlength{\aboverulesep}{0pt}
\setlength{\belowrulesep}{0pt}
\newcommand*{\rulestrut}{\rule{0pt}{\dimexpr\ht\strutbox+\belowsep\relax}}
{\footnotesize
\begin{tabular}{@{}l
  >{\columncolor{gray!15}[\tabcolsep][3pt]}r@{\hspace{3pt}}%
  >{\columncolor{gray!15}[1.5pt][\tabcolsep]}r
  r@{/}l|
  >{\columncolor{gray!15}[\tabcolsep][3pt]}r@{\hspace{3pt}}%
  >{\columncolor{gray!15}[1.5pt][\tabcolsep]}r
  r r@{/}l@{}}
\toprule
\rulestrut & \multicolumn{4}{c|}{\whole} & \multicolumn{5}{c@{}}{\sys} \\[\abovesep]
\cmidrule{2-5}\cmidrule{6-10}
\rulestrut Candidates
& \multicolumn{2}{>{\columncolor{gray!15}}c}{Accuracy}
& \multicolumn{2}{c|}{Time (s)}
& \multicolumn{2}{>{\columncolor{gray!15}}c}{Accuracy}
& Located & \multicolumn{2}{c@{}}{Time (s)} \\[\abovesep]
\midrule
\rulestrut \emph{Task 1} & & & \multicolumn{2}{c|}{} & & & & \multicolumn{2}{c@{}}{} \\
\quad correct   & 3/3\phantom{0} & & 1418 & 1443 & 3/3\phantom{0} & & \multicolumn{1}{c}{--} & 1344 & 1232 \\
\quad incorrect & {\bf 2/6}\phantom{0} & &  649 &  520 & {\bf 5/6}\phantom{0} & & {\bf 4/6}\phantom{0} &  950 &  975 \\[3pt]
\emph{Task 2} & & & \multicolumn{2}{c|}{} & & & & \multicolumn{2}{c@{}}{} \\
\quad correct   & 3/3\phantom{0} & &  694 &  530 & 3/3\phantom{0} & & \multicolumn{1}{c}{--} &  528 &  548 \\
\quad incorrect & {\bf 2/5}\phantom{0} & &  339 &  258 & {\bf 5/5}\phantom{0} & & \bf{5/5}\phantom{0} &  834 &  564 \\[\abovesep]
\midrule
\rulestrut\emph{Both tasks} & & & \multicolumn{2}{c|}{} & & & & \multicolumn{2}{c@{}}{} \\
\quad correct   & 6/6\phantom{0} & (100\%) & 1056 &  947 & 6/6\phantom{0} & (100\%) & \multicolumn{1}{c}{--} &  936 &  708 \\
\quad incorrect & {\bf 4/11} & (36\%) &  508 &  486 & {\bf 10/11} & (91\%) & {\bf 9/11} &  897 &  931 \\
\quad all       & {\bf 10/17} & (59\%) &  701 &  530 & {\bf 16/17} & (94\%) & \multicolumn{1}{c}{--} &  911 &  823 \\[\abovesep]
\bottomrule
\end{tabular}
}
\end{table}

We test \sys{} with human validators and find validation on decomposition more effective
for them than on the whole query: participants catch errors that they miss in whole queries,
without rejecting correct candidates, and their rejections point to the
errors.

\emph{Study Design.}
Our Campus Institutional Review Board (IRB) rules preclude a study within an ongoing database course or involving students advised by faculty, so we
invited 218 SQL-proficient students who had taken an
undergraduate or graduate database course earlier (and offered a small  gift card for their time); 20 of them volunteered.
Participants took part in person, except a few who were not local and joined remotely, and were monitored over Zoom. None of them consulted an LLM. They validated
candidate queries for two questions from BEAVER's dw split, Task~1 and Task~2,
both marked as needing no domain knowledge. Each question had four candidate
queries, two correct and two incorrect; we dropped one correct query per
question because it received too few validations. Each participant performed a reference-free validation of one candidate of each
task, Task~1 first, with the presentation counterbalanced: half of the
participants validated Task~1 as a whole query and Task~2 decomposed, the other
half the reverse.
Candidates were assigned by balanced randomization, so every
candidate was validated equally often both ways, giving 17 validations each way. 
A whole-query validation task shows the question, the schema, and
the SQL. A decomposed task shows the blocks of \sys{}'s decomposition, each
with its requirement, the descriptions of its input blocks, the schema, and its SQL,
and asks for each block whether its SQL does what its requirement asks; a
candidate is rejected if any of its blocks is. We record each verdict, its
explanation, and the time spent on each page.
Appendix~\ref{app:userstudy} gives more details on the questions. 

\emph{Effect on Accuracy.}
Decomposition helps people catch errors. With \sys{}, participants
rejected the incorrect candidates in 91\% of the validations (10 of 11),
against 36\% (4 of 11) with whole queries (Table~\ref{tab:userstudy}). On
correct candidates, the two ways performed alike: participants accepted the
candidate in all six validations with each. Overall, accuracy rose from 59\%
to 94\%. 
This agrees with the LLM
experiments (Section~\ref{subsec:exp-effectiveness}), with one difference: an
LLM hallucinates defects with some probability and rejected two more correct
candidates with \sys{}, whereas decomposition did not make human
verdicts on correct candidates worse. The gain holds for both tasks. Task~1 is
the harder validation task: with whole
queries, only 2 of 6 validations of its incorrect candidates caught the error,
and with \sys{}, 5 of 6. Task~2 is easier, yet whole-query validation still
missed the error in 3 of 5 validations, while \sys{} caught all five.

\emph{Effect on Validation Time.}
With \sys{}, the time per block is stable: a block took a median of
233\,s in Task~1 and 225\,s in Task~2. The time of a whole query follows the task: a median of 726\,s in Task~1, the
harder task, in which participants also had to learn the schema that both
tasks share, and 391\,s in Task~2.
On correct
candidates, \sys{} took no longer than whole-query validation (708 vs.\
947\,s). On incorrect candidates, \sys{} took longer (931
vs.\ 486\,s). A likely reason is that the participants who found the error in
a whole query could make their judgment as soon as they saw the
error and skipped the rest of the query, whereas with \sys{}, a participant who
finds an error in an early block still validates the later ones. In practice, a validator can stop at the first rejected block, since
one rejection rejects the candidate.

\emph{Localization of Errors.}
Decomposition also helps locate errors. Of the 10 rejections with
\sys{}, 9 fell on the defective block or on a block that reads its result
(Table~\ref{tab:userstudy}), whereas a whole-query verdict carries no location. 
Participants were proficient in SQL: in 12 of their 14
rejections of incorrect candidates, the explanation named the defect, such as a \sqlkw{LIMIT~1} that keeps only one of several tied rows, a \sqlkw{CASE}
branch that is never reached because an earlier condition covers it, or an
integer division that truncates an average.

\section{Experiments with LLM Validators}
\label{sec:evaluation}
We evaluate whether decomposition helps validators judge candidate queries,
and whether \sys{} chooses decompositions that suit them. LLM validators let us
evaluate at scale 
by judging 305 candidate queries that a text-to-SQL
system generates for questions from BEAVER~\cite{chen2026beaver} and Spider~2.0~\cite{lei2024spider}, each under several
decompositions and granularities,
to answer three research questions:

\textbf{(1) RQ1 (Comparison with Validation Baselines).} Does \sys{} catch
errors that baseline validation methods miss with the same LLM, and on which
queries does it help most?

\textbf{(2) RQ2 (Comparison of Decompositions).} Do the
decompositions that \sys{} chooses yield more accurate validation and simpler
blocks than syntactic or random decompositions?

\textbf{(3) RQ3 (Effect of Granularity across Validators).} How does
decomposition granularity affect different LLM validators?

\subsection{Experimental Setup}
\label{subsec:setup}

All experiments ran on a Linux server with an AMD EPYC 7R13 processor (48 cores,
2.6\,GHz) and 256\,GB of RAM. Up to
8 candidates are processed in parallel,
each with at most
4 concurrent LLM calls.

\begin{table}[t]
\centering
\footnotesize
\caption{The 305 candidate queries: mean (maximum).}
\label{tab:benchmark}
\setlength{\tabcolsep}{4pt}
\begin{tabular}{@{}lrrr@{}}
\toprule
& BEAVER & Spider~2.0 & All \\
\midrule
Questions & 270 & 35 & 305 \\
Correct / incorrect candidates & 46 / 224 & 16 / 19 & 62 / 243 \\
\midrule
Tokens & 210 (752) & 206 (796) & 209 (796) \\
Relations & 3.2 (6) & 3.0 (7) & 3.2 (7) \\
Joins & 2.7 (12) & 3.0 (9) & 2.8 (12) \\
Predicates & 6.4 (18) & 6.1 (26) & 6.3 (26) \\
Aggregations & 1.8 (9) & 1.7 (6) & 1.8 (9) \\
Subqueries & 0.8 (7) & 0.3 (3) & 0.8 (7) \\
CTEs & 2.6 (13) & 2.7 (10) & 2.6 (13) \\
\bottomrule
\end{tabular}
\end{table}

\emph{Benchmarks.}
Our benchmark has 305 questions. Of these, 270 come from the Neutron split of
BEAVER~\cite{chen2026beaver}, 1,017 queries over an enterprise database with 175
tables. BEAVER labels each query by its structure and by whether it requires
domain knowledge; we draw 30 questions at random from each of the nine
resulting categories, among those whose gold query \sys{} supports (Section~\ref{sec:implementation}). The
other 35 are the Spider~2.0~\cite{lei2024spider} tasks on local databases whose
gold query \sys{} supports. For each question, DIN-SQL~\cite{pourreza2023dinsql}
with GPT-6 Luna~\cite{openai2026models} at maximal reasoning effort generates
one candidate query. The candidates are long and layered
(Table~\ref{tab:benchmark}): 209 tokens, 3.2 base tables, and 2.6 CTEs on
average. 

\emph{Labels.}
A candidate is \emph{correct} if it passes its benchmark's official evaluation:
for BEAVER, it returns the same set of rows as the gold query; for Spider~2.0,
the benchmark's evaluator accepts its result. Of the 305 candidates, 62 are
correct and 243 incorrect (BEAVER 46 and 224, Spider~2.0 16 and 19). 

\emph{Validators and configuration.}
Three LLMs serve as validators: GPT-5.5, GPT-5.6 Sol, and GPT-6 Astra~\cite{openai2026models}, each at
medium reasoning effort.
Every LLM call runs through the Codex CLI~\cite{openai2025codex} in a
sandbox that holds only its prompt, so a validator sees only its task, never the gold query, the labels, or a previous
result.
The reliability model uses $\beta = \ln 2$, $\gamma = 0.25$, and
$\mu = 10^{-3}$, with default thresholds $\tau^*$: 2 for
\textsf{inner\_join}, \textsf{bool\_mix}, and \textsf{order\_limit}, 3 for
\textsf{pred}, 4 for \textsf{expr} and \textsf{n\_relations}, 5 for
\textsf{n\_optypes}, 150 tokens for \textsf{length}, and 1 for every other
feature (Table~\ref{tab:features}). We set $\tau^*$ before the user study
from the constructs on which people and LLMs most often err
(Section~\ref{subsec:sources}) and our own judgment of SQL difficulty.
GPT-6 Luna writes requirements and input
descriptions at maximal reasoning effort and rates coherence and relevance at
high effort. The time budget of Section~\ref{sec:implementation} is one
hour per candidate.
To vary granularity, we scale every threshold by
a factor $\alpha$, $\tau = \alpha\tau^*$: a larger $\alpha$ models a validator
with a wider range and yields coarser decompositions.
Unless stated otherwise, the validator is GPT-5.6 Sol and $\alpha = 1$.

\emph{Validation baselines.}
\textbf{1. \whole} validates the candidate as a whole against the question, without
decomposing it. \wholek runs \whole $k$ times, where $k$ is the number of
distinct blocks in \sys{}'s decomposition of the candidate, and rejects the
candidate if any run rejects it. We also run two validation methods of Alrashed
et al.~\cite{alrashed2026verification}. \textbf{2. \emph{Round-trip critique} (\rt)} has the
LLM translate the candidate back into a question without seeing the original
question, and accepts the candidate if the LLM gives the two questions a
similarity score of at least 4 out of 5. \textbf{3. \emph{Synthetic execution} (\se)} has
the LLM write the expected answer to the question and then a database instance
that yields it, executes the candidate on that instance, and accepts the
candidate if its result equals the expected answer. \sys{}~+~\rt and
\sys{}~+~\se apply \rt and \se to each of \sys{}'s blocks instead of the whole
candidate, with the block's requirement as the question and its inputs as
tables; a block that \se cannot execute keeps \sys{}'s verdict. 

\emph{Decomposition strategies.}
We compare \sys{} with three other decomposition strategies: \syntactic cuts at every CTE,
derived table, and subquery written in the candidate where the cut is legal;
\randomk randomly samples a legal decomposition with as many distinct blocks as
\sys{}'s; and \sys{} with $\gamma = 0$ optimizes the structural risk model
alone. All decompositions keep the forced cuts of Section~\ref{subsec:blocks} and
share task assignment, presentation, and verdict composition, so they differ
only in their blocks.

\emph{Metrics.}
We take an incorrect candidate as the positive class, so a true positive is an error that the validator detects (Appendix~\ref{app:metrics}):
\begin{center}
\footnotesize
\begin{tabular}{@{}lcc@{}}
\toprule
& Validator rejects & Validator accepts \\
\midrule
Incorrect candidate & TP & FN \\
Correct candidate & FP & TN \\
\bottomrule
\end{tabular}
\end{center}
We report accuracy, precision, recall (the share of incorrect candidates
rejected), specificity (the share of correct candidates accepted), and F1.
No method judges the five candidates that \sys{} does not support
(four use \sqlkw{LATERAL}, one a table function). A candidate without a
verdict stays in the denominators but in no numerator.

\emph{Token cost.}
\label{subsec:exp-cost}%
We report medians per candidate, with GPT-5.6 Sol as the validator. \sys{} makes 9 LLM calls: two for semantic scores, four to write requirements and input descriptions, and
three validator calls.
At list prices~\cite{openai2026models}, validating a query costs \$0.08 using \sys{}, against \$0.02 for \whole. Of \sys{}'s cost, \$0.01 goes to the
semantic scores, requirements, and input descriptions, all computed once per question--query pair
and shared by every validator.
Runtimes are not comparable across methods
because Codex's rate limits throttled our concurrent calls.

\subsection{RQ1: Comparison with Validation Baselines}
\label{subsec:exp-effectiveness}

\begin{table}[t]
\centering
\footnotesize
\caption{Validation outcomes on 305 candidates.}
\label{tab:main}
\setlength{\tabcolsep}{2.5pt}
\begin{tabular}{@{}l rrrr@{\hspace{8pt}}rrrrr@{}}
\toprule
& \multicolumn{4}{c}{Candidates} & \multicolumn{5}{c}{Measures (\%)} \\
\cmidrule(lr){2-5}\cmidrule(l){6-10}
Method & TP & FN & FP & TN & Acc & Prec & Rec & Spec & F1 \\
\midrule
\multicolumn{10}{@{}l}{\emph{Without execution}} \\
\quad \whole           &  61 & 182 &  6 & 55 & 38.0 & 91.0 & 25.1 & \textbf{88.7} & 39.4 \\
\quad \wholek          &  83 & 160 &  6 & 55 & 45.2 & \textbf{93.3} & 34.2 & \textbf{88.7} & 50.0 \\
\quad \sys{}           &  97 & 146 &  8 & 53 & \textbf{49.2} & 92.4 & \textbf{39.9} & 85.5 & \textbf{55.7} \\
\cmidrule(l{1em}){1-10}
\quad \rt              &  43 & 200 &  3 & 58 & 33.1 & \textbf{93.5} & 17.7 & \textbf{93.5} & 29.8 \\
\quad \sys{} + \rt     & 153 &  90 & 26 & 35 & \textbf{61.6} & 85.5 & \textbf{63.0} & 56.5 & \textbf{72.5} \\
\midrule
\multicolumn{10}{@{}l}{\emph{With execution}} \\
\quad \se              & 188 &  55 & 33 & 27 & 70.5 & \textbf{85.1} & 77.4 & \textbf{43.5} & 81.0 \\
\quad \sys{} + \se     & 220 &  23 & 46 & 15 & \textbf{77.0} & 82.7 & \textbf{90.5} & 24.2 & \textbf{86.4} \\
\bottomrule
\end{tabular}
\end{table}

\begin{figure}[t]
\centering
\includegraphics[width=\columnwidth]{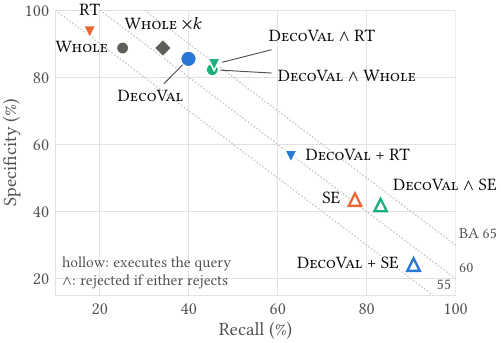}
\caption{Recall and specificity of the
methods of Table~\ref{tab:main} and their combinations with \sys{}.
Dotted lines mark equal balanced accuracy (BA), the mean of recall and
specificity.}
\label{fig:tradeoff}
\end{figure}

\emph{Decomposition vs.\ whole queries.}
With the same LLM validator, \sys{} catches many errors that validating the
whole query misses (Table~\ref{tab:main}, Figure~\ref{fig:tradeoff}): it
rejects 97 of the 243 incorrect candidates against 61 for \whole, raising
recall from 25.1\% to 39.9\% and F1 from 39.4 to 55.7 with two more false
rejections. Decomposition contributes detections beyond those of repeated
judgments: \wholek, which makes as many judgments as \sys{} and composes them
with the same rule, reaches a recall of 34.2\% and an F1 of 50.0, yet still finds 14 fewer errors than \sys{}.

Decomposition also raises the recall of the two methods of Alrashed et al.,
which err in opposite directions on whole queries. \rt rarely rejects because
comparing two natural-language questions misses many differences that change
the result: it catches 43 incorrect candidates and gives 89 of the 200 it
misses a similarity score of 5. \se catches 188 incorrect candidates but
rejects 33 of the 62 correct ones. On \sys{}'s blocks, \rt catches 153
incorrect candidates and \se 220, or 90.5\% of all incorrect queries, while their false rejections
rise from 3 to 26 and from 33 to 46. Decomposition thus raises the BA of the LLM validator and
\rt, from 56.9 to 62.7 and from 55.6 to 59.7, and lowers that of \se, from
60.5 to 57.4.

\begin{figure}[t]
\centering
\includegraphics[width=\columnwidth]{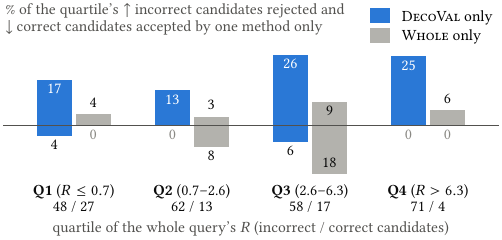}
\caption{Where \sys{} and \whole disagree, 
by quartile of the whole query's structural risk $R$.}
\label{fig:length}
\end{figure}

\emph{Where decomposition helps most.}
Decomposition helps most on complex queries, as the reliability model
predicts. Figure~\ref{fig:length} groups the candidates by the structural risk
$R$ of the whole query (Section~\ref{subsec:reliability}), which grows as the
query's features exceed the validator's thresholds. The recall of \sys{}
exceeds that of \whole by 12.5 and 9.7 points in the lower two quartiles and
by 17.2 and 19.7 points in the upper two, where \sys{} alone rejects 33
incorrect candidates and \whole alone 9. In every quartile, \sys{} has at most
two more false rejections than \whole.
Blocks and whole queries complement each other. \whole catches 13 errors that
\sys{} misses, some of which concern how the steps fit together and are easier
to see on the whole query. In one BEAVER candidate, a left join keeps unmatched
rows; the aggregates skip their \textsc{NULL} values, but a later grouping
turns each of them into an extra row of the result. \sys{} puts the join and
the grouping in separate blocks with correct requirements, and each looks right
on its own; \whole sees both together and rejects the candidate. Combining
verdicts by conjunction ($\wedge$) catches more errors: \sys{}~$\wedge$~\whole
rejects 110 incorrect candidates and \sys{}~$\wedge$~\rt 111, against 97 for
\sys{}, with one or two more false rejections; \sys{}~$\wedge$~\rt reaches the
highest BA (64.8) in Figure~\ref{fig:tradeoff}.

\subsection{RQ2: Comparison of Decompositions}
\label{subsec:exp-strategies}

\begin{table}[t]
\centering
\footnotesize
\caption{
Decomposition strategies.
\#~blocks: mean distinct blocks per candidate; Out of range:
blocks with a feature over its threshold in~$\tau^*$; $R$: mean structural risk
of a block.}
\label{tab:strategies}
\setlength{\tabcolsep}{3pt}
\begin{tabular}{@{}l rrr@{\hspace{8pt}}rrr@{}}
\toprule
& \multicolumn{3}{c}{Decomposition} & \multicolumn{3}{c}{Measures (\%)} \\
\cmidrule(lr){2-4}\cmidrule(l){5-7}
Strategy & \#~blocks & Out of range (\%) & $R$ & Rec & Spec & BA \\
\midrule
\whole               & 1.0 & 92.0 & 4.34 & 25.1 & 88.7 & 56.9 \\
\midrule
\syntactic           & 4.4 & 22.9 & 0.27 & 42.0 & 74.2 & 58.1 \\
\randomk             & 4.0 & 30.6 & 0.34 & 39.1 & 83.9 & 61.5 \\
\sys{}, $\gamma = 0$ & 3.9 & \textbf{13.1} & \textbf{0.13} & \textbf{42.4} & 80.6 & 61.5 \\
\sys{}               & 4.0 & 13.4 & \textbf{0.13} & 39.9 & \textbf{85.5} & \textbf{62.7} \\
\bottomrule
\end{tabular}
\end{table}

\sys{} produces the fewest and simplest blocks and the best validation
(Table~\ref{tab:strategies}). It cuts a candidate into 3.9 to 4.0 blocks on
average, against 4.4 for \syntactic, and under the reliability model it
optimizes, its blocks carry less than half the structural risk of those of
\syntactic and \randomk. With $\gamma = 0$ it catches the most errors, and by
default it reaches the highest BA, 62.7. \syntactic depends on how the query is
written. DIN-SQL writes 238 of the 305 candidates as pipelines of CTEs, which
give \syntactic natural places to cut; on the other 67 it finds few, and its
blocks carry six times the structural risk of \sys{}'s.

\randomk leaves the difficulty of its blocks to chance: with as many blocks as
\sys{}, its riskiest block carries more than five times the structural risk of
\sys{}'s on the tenth of the candidates where the two differ most. The
structural risk alone accounts for \sys{}'s advantage: without the semantic
score ($\gamma = 0$), its blocks are as simple, and the semantic score trades
6 detections for 3 fewer false rejections.

\subsection{RQ3: Effect of Granularity across Validators}
\label{subsec:exp-granularity}

\begin{figure}[t]
\centering
\includegraphics[width=\columnwidth]{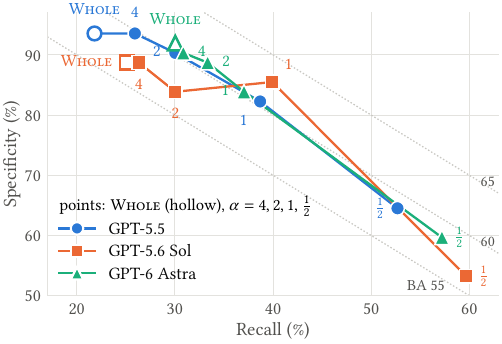}
\caption{Recall and specificity of three LLM
validators from the whole query to $\alpha = \frac{1}{2}$.
Dotted lines join equal BA.
}
\label{fig:granularity}
\end{figure}

Granularity trades recall against specificity for every validator
(Figure~\ref{fig:granularity}). \sys{} rejects a candidate as soon as it
rejects one block, so every additional judgment is another chance to reject.
From \whole to $\alpha = \frac{1}{2}$, where a candidate has 7.9 blocks on
average, each validator gains 27 to 35 points of recall and loses 29 to 35
points of specificity. Conversely, a larger $\alpha$ makes fewer judgments and
raises specificity toward that of \whole.

Recall and specificity move along a line of nearly equal BA for GPT-5.5 and
GPT-6 Astra, whose BA varies by less than 3 points across granularities;
GPT-5.6 Sol has a clear optimum at $\alpha = 1$, 5.8 points above \whole and at
least 5.2 above any other granularity.
The choice of granularity
thus depends on the validator.

\section{Conclusion And Future Work}
\label{sec:conclusion}

Validation-aware query decomposition rewrites a candidate query into simpler
blocks that recompose to it by construction and chooses the decomposition that a
given validator is expected to judge most reliably. It more than doubled the
share of errors that participants in our user study caught, from 36\% to 91\%,
and raised the share that three LLM validators caught by 7.0--16.9 points.
Future work can improve three parts of \sys{}. Its rewrites only reorder joins
and move filters, which keeps blocks close to the candidate's SQL; whether
decorrelation and other rewrites help validators is open. Its reliability model
uses default thresholds and could be fitted to a validator's past verdicts. Its
block requirements and descriptions come from LLMs; finding more reliable
ways to assign tasks to blocks is also an open problem.

\bibliographystyle{ACM-Reference-Format}
\bibliography{bib}

\clearpage
\appendix
\section{Proofs}
\label{app:proofs}

\theoremstyle{acmplain}
\newcommand*{\restatedname}{}
\newtheorem*{restatedresult}{\restatedname}
\newenvironment{restated}[3]{%
  \renewcommand*{\restatedname}{#1~\ref{#2}}%
  \begin{restatedresult}[#3]}{\end{restatedresult}}

\subsection{Exactness without Shareable Classes}
\label{app:proof-dp}

\begin{restated}{Theorem}{thm:dp}{Exactness without shareable classes}
If no class is shareable, DP returns a decomposition in~$\Dspace$ of
minimum~$J_\theta$.
\end{restated}

\emph{Proof outline.}
DP picks a block for every group bottom-up, combining the best choices for
the inputs of a block (Equation~\eqref{eq:dp}). This is the dynamic
programming of memo-based plan
enumeration~\cite{selinger1979access,graefe1995cascades}, and it is exact
under the two conditions of the principle of optimality: the cost of a
decomposition is the sum of the costs of its blocks, and the inputs of a
block can be decomposed independently of each other. The first condition
holds as no class is shareable, so that every block is a task of its own.
The second is what needs proof in our setting: Lemma~\ref{lem:tree}
shows it, and Lemma~\ref{lem:dp} then derives the recurrence of DP by the
standard argument.

\emph{Scope.}
Definition~\ref{def:problem} takes a finite, nonempty candidate set. As the
memo is finite, so is~$\Dspace$; here and in Appendix~\ref{app:proof-astar},
we consider queries for which it is nonempty.
(A query that only scans a table and selects columns has no block under
rule~(1) of Section~\ref{subsec:blocks}.)

\emph{Notation.}
Let $\mathcal{H}$ consist of~$g_0$ and the forced cuts, which DP solves
separately (Section~\ref{subsec:dp}). For a block $b=(g,I)$, let
$\mathit{out}(b)=g$, let $I^{\circ}(b)\subseteq I$ be its \emph{non-forced
inputs}, those that are not forced cuts, and let $w(b)=w_\theta(c)$ for its
class~$c$. A class is shareable if it contains two or more blocks
of~$\mathcal{B}$; every other class contains exactly one. For a set~$D$ of
blocks, $\mathit{out}(D)$ and $\classes(D)$ denote the output groups and the
classes of its blocks, and $\omega(D)=\sum_{b\in D}\omega(b)$ for a
cost~$\omega$ on blocks. An input of a block is the group of a proper
subplan of a plan of its output, so it is a proper descendant of the output
in the memo; as the memo is acyclic, we can define sets for groups and prove
claims about them bottom-up, from descendants to ancestors.

\emph{Sub-decompositions.}
A \emph{sub-decomposition} of a group~$p$ decomposes the part of~$q$ that $p$
computes: it consists of a block~$b$ with output~$p$ and a sub-decomposition
of each non-forced input of~$b$. Bottom-up, the set $\mathcal{D}(p)$ of
sub-decompositions of~$p$ is thus
\[
\mathcal{D}(p)=\Big\{\{b\}\cup\textstyle\bigcup_{i\in I^{\circ}(b)}D_i \;:\;
b\in\mathcal{B},\ \mathit{out}(b)=p,\ D_i\in\mathcal{D}(i)\Big\}.
\]
Let $\mathit{Out}(p)=\bigcup_{D\in\mathcal{D}(p)}\mathit{out}(D)$ be the set
of groups that receive a block in some sub-decomposition of~$p$.

\begin{lemma}[Independence]
\label{lem:tree}
\textup{(a)}~For every block $b\in\mathcal{B}$, the sets $\mathit{Out}(i)$,
$i\in I^{\circ}(b)$, are pairwise disjoint and do not contain
$\mathit{out}(b)$, and the sets $\mathit{Out}(r)$, $r\in\mathcal{H}$, are
pairwise disjoint.
\textup{(b)}~Hence the blocks of every $D\in\mathcal{D}(p)$ have distinct
output groups, and if $D=\{b\}\cup\bigcup_{i\in I^{\circ}(b)}D_i$, then
$\omega(D)=\omega(b)+\sum_{i\in I^{\circ}(b)}\omega(D_i)$ for every
cost~$\omega$.
\textup{(c)}~$\Dspace$ consists of the unions $\bigcup_{r\in\mathcal{H}}D_r$
with $D_r\in\mathcal{D}(r)$, whose sets~$D_r$ are pairwise disjoint.
\end{lemma}

\noindent Part~(a) says that the non-forced inputs of a block, and the groups
of~$\mathcal{H}$, can be decomposed separately: no block can serve two of
them.

\begin{proof}
(a)~We assign to each group the occurrences, that is, the FROM items
(Section~\ref{subsec:plans}), that its plans read; all plans of a group read
the same ones (Section~\ref{subsec:memo}). Two conventions complete this
assignment. First, a \sqlkw{SELECT} without \sqlkw{FROM}, such as
\sqlkw{SELECT 1 + 1}, reads a one-row source, which counts as an occurrence
of its own; so two equal such subqueries at different positions are
different groups. Second, a reference to a CTE that $q$ reads from several
places counts as an occurrence, but the occurrences in the CTE's body, which
all its references share, do not; they count only for the groups
within the body. This ignores no block, as no block reaches
into such a body from outside: the body is a forced cut, which every block
that reads it takes as an input (rule~(3) of Section~\ref{subsec:blocks}),
unless it cannot be a block, and then no group in it outside the
forced cuts that it reads is the output of a block (rule~(1)). Three facts
follow:
\begin{enumerate}[label=(\roman*),leftmargin=*,nosep]
\item every group reads at least one occurrence, at a leaf of its plans or as
a reference above that leaf;
\item every occurrence counted for a plan lies at one position of it, so the
non-forced inputs of~$b$, which are disjoint parts of a plan of
$\mathit{out}(b)$, read different occurrences;
\item a group in $\mathit{Out}(i)$ lies within the part of~$i$, so it reads
only occurrences that $i$ reads.
\end{enumerate}
By (i) and~(iii), a group in two of the sets $\mathit{Out}(i)$ would read an
occurrence that two inputs read, which (ii) excludes. Moreover, every group
in $\mathit{Out}(i)$ is a proper descendant of $\mathit{out}(b)$, so none is
$\mathit{out}(b)$. The same argument applies to~$g_0$ and the forced cuts,
which read each other only through references.

(b)~By bottom-up induction, the blocks of each~$D_i$ in
$D=\{b\}\cup\bigcup_{i}D_i$ have distinct output groups. These lie in
$\mathit{Out}(i)$, and by~(a), these sets are pairwise disjoint and do not
contain $\mathit{out}(b)$. So all blocks of~$D$ have distinct output groups,
the union is disjoint, and $\omega(D)=\omega(b)+\sum_{i}\omega(D_i)$.

(c)~Every $D\in\Dspace$ splits as stated. It has a block for~$g_0$ and for
every forced cut and, for each of its blocks, a block for every non-forced
input, and no other blocks, as all its blocks are reachable from the root
(Section~\ref{subsec:blocks}). Following the non-forced inputs from the block
of each $r\in\mathcal{H}$ thus collects a set $D_r\in\mathcal{D}(r)$, and
these sets cover~$D$ and are disjoint by (a) and~(b). Conversely, by (a)
and~(b), a union of sets $D_r\in\mathcal{D}(r)$, $r\in\mathcal{H}$, has
exactly one block for each group that it decomposes, including every input
of its blocks, and it satisfies the rules of Section~\ref{subsec:blocks},
which restrict each block separately; so it lies in~$\Dspace$.
\end{proof}

\noindent As $\mathcal{B}$ consists of the blocks of these decompositions,
$\mathcal{D}(\mathit{out}(b))$ and $\mathcal{D}(i)$, $i\in I^{\circ}(b)$, are
nonempty for every $b\in\mathcal{B}$. The next lemma is the standard
optimal-substructure argument of dynamic programming; we state it for any
cost~$\omega$, as Appendix~\ref{app:proof-astar} also applies it to other
costs.

\begin{lemma}[Recurrence]
\label{lem:dp}
For a cost~$\omega$ on blocks and a group~$p$ that is the output of a block
of~$\mathcal{B}$, let $\mathit{best}_\omega(p)$ be the least cost $\omega(D)$
of a sub-decomposition $D\in\mathcal{D}(p)$. Then
\[
\mathit{best}_\omega(p)=\min_{b\in\mathcal{B},\;\mathit{out}(b)=p}
\Big[\omega(b)+\sum_{i\in I^{\circ}(b)}\mathit{best}_\omega(i)\Big],
\]
which is Equation~\eqref{eq:dp} for the cost~$\omega$ with the forced cuts
left out (Section~\ref{subsec:dp}). Choosing a minimizing block at~$p$ and,
recursively, at its non-forced inputs yields a sub-decomposition
$D^{*}_\omega(p)\in\mathcal{D}(p)$ of cost $\mathit{best}_\omega(p)$.
\end{lemma}

\begin{proof}
A sub-decomposition of~$p$ consists of a block~$b$ with output~$p$ and
sub-decompositions $D_i\in\mathcal{D}(i)$ of its non-forced inputs, chosen
independently of each other, and its cost is
$\omega(b)+\sum_{i}\omega(D_i)$ by Lemma~\ref{lem:tree}(b). For a
fixed~$b$, choosing each~$D_i$ of least cost minimizes this sum, to
$\omega(b)+\sum_{i}\mathit{best}_\omega(i)$; minimizing over~$b$ gives the
recurrence, and these choices form $D^{*}_\omega(p)$.
\end{proof}

\begin{proof}[Proof of Theorem~\ref{thm:dp}]
As no class is shareable, distinct blocks have distinct classes, so
$J_\theta(D)=w(D)$ for every set~$D$ of blocks. The sets
$\mathcal{D}(r)$, $r\in\mathcal{H}$, are nonempty, as $\Dspace$ is, and DP
returns $D^{*}=\bigcup_{r\in\mathcal{H}}D^{*}_w(r)$ (Lemma~\ref{lem:dp}),
which lies in~$\Dspace$ by Lemma~\ref{lem:tree}(c). By the same lemma, every
$D\in\Dspace$ is the disjoint union of sets $D_r\in\mathcal{D}(r)$,
$r\in\mathcal{H}$, and so is~$D^{*}$, of the sets $D^{*}_w(r)$; hence
\[
J_\theta(D)=\textstyle\sum_{r\in\mathcal{H}}w(D_r)
\ge\sum_{r\in\mathcal{H}}\mathit{best}_w(r)=J_\theta(D^{*}).\qedhere
\]
\end{proof}

\subsection{Exactness of the Search}
\label{app:proof-astar}

\begin{restated}{Theorem}{thm:astar}{Exactness of the search}
The heuristic~$h$ is admissible and consistent, and
Algorithm~\ref{alg:search} returns a decomposition in~$\Dspace$ of
minimum~$J_\theta$.
\end{restated}

\emph{Proof outline.}
A decomposition can share work in two ways. A group read from several
places, the body of a CTE that $q$ reads more than once, is a forced cut,
which DP solves once (Appendix~\ref{app:proof-dp}). Different blocks can
share one task, by belonging to one shareable class; this couples their
costs, and the search therefore records the shareable classes paid so far
in~$\Phi$. Without shareable classes, Algorithm~\ref{alg:search} returns the
result of DP, which is optimal by Theorem~\ref{thm:dp}. Otherwise, it runs A*
on a graph of search states, and A* returns a cheapest path to a goal when its
heuristic is consistent (Lemma~\ref{lem:astar}). Three facts about the search
then give the theorem: every goal yields a decomposition whose
cost~$J_\theta$ is what the search paid to reach it
(Lemma~\ref{lem:invariants}); \textsc{Fold} preserves the optimal
cost, as every decomposition has a reachable goal that is at least as cheap
(Lemma~\ref{lem:complete}); and $h$ is consistent (Lemma~\ref{lem:heuristic}).

\emph{Notation.}
We keep the scope, the notation, and the results of
Appendix~\ref{app:proof-dp}. Every class cost is nonnegative, as $\mu\ge0$,
$R(\theta,c)\ge0$, $\gamma\ge0$, and $s(c)\in(0,1]$. A group~$p$
\emph{can use} a class if the class occurs in some sub-decomposition of~$p$.
If a block with output~$p$ has a non-forced input~$i$, every
sub-decomposition of~$i$ is contained in some sub-decomposition of~$p$; so
$\mathit{Out}(i)\subseteq\mathit{Out}(p)$, and $i$ can use only classes that
$p$ can use. For a set $\Phi\subseteq\classes_{\mathit{sh}}$ of paid
classes, $w_\Phi$ is the cost~$w$ with the classes of~$\Phi$ free:
$w_\Phi(b)=0$ if the class of~$b$ lies in~$\Phi$, and $w_\Phi(b)=w(b)$
otherwise. The heuristic makes every shareable class free: with
$\hat w=w_{\classes_{\mathit{sh}}}$, it is
$h(\sigma)=\sum_{p\in\Pi}\mathit{best}_{\hat w}(p)$ for a state
$\sigma=(\Pi,\Phi)$.

\emph{The search graph.}
A state $\sigma=(\Pi,\Phi)$ holds the pending groups~$\Pi$ and the
paid shareable classes~$\Phi$; along with it, Algorithm~\ref{alg:search}
keeps the set~$D_\sigma$ of blocks chosen so far and the part~$J(\sigma)$
of~$J_\theta$ paid so far. Three kinds of steps change them:
\begin{itemize}[leftmargin=2.4em,labelsep=0.4em,nosep]
\item[(M)] A \emph{move} gives the first pending group~$p$ a block~$b$ with
output~$p$. It pays $w_\Phi(b)$, adds~$b$ to~$D_\sigma$, replaces~$p$ by the
non-forced inputs of~$b$ in~$\Pi$, and adds the class of~$b$ to~$\Phi$ if that
class is shareable.
\item[(F1)] \textsc{Fold} solves a pending group~$p$ that can use no
shareable class outside~$\Phi$. It pays $\mathit{best}_{w_\Phi}(p)$, adds
$D^{*}_{w_\Phi}(p)$ to~$D_\sigma$, and removes~$p$ from~$\Pi$.
\item[(F2)] \textsc{Fold} drops from~$\Phi$ the classes that no
pending group can use.
\end{itemize}
The search begins at $(\mathcal{H},\emptyset)$ with $D_\sigma=\emptyset$ and
$J(\sigma)=0$, and \textsc{Fold} turns this into the start state. An edge
from~$\sigma$ is a move followed by \textsc{Fold}, and its cost is the amount
they pay; the edges of a state and their costs depend only on $(\Pi,\Phi)$.
The goals are the states with $\Pi=\emptyset$. Algorithm~\ref{alg:search} runs
A* on this graph and keeps only the cheapest path found to each state
(Section~\ref{subsec:astar}): line~9 relaxes $\textsc{Fold}(\sigma')$, that is,
the state enters~$Q$, with $D_{\sigma'}$ and~$J(\sigma')$, only if no path to
it found before is as cheap, and it then replaces the entry of such a path.

\emph{Step~(F1).}
It is the DP of Appendix~\ref{app:proof-dp} applied to~$p$, with
the classes of~$\Phi$ free, and it is exact for the same reason as
Theorem~\ref{thm:dp}. As $p$ can use no shareable class outside~$\Phi$, every
class of a sub-decomposition $D_p\in\mathcal{D}(p)$ outside~$\Phi$ is not
shareable and has its only block in~$D_p$; so
\[
w_\Phi(D_p)=\textstyle\sum_{c\in\classes(D_p)\setminus\Phi}w_\theta(c),
\tag{$\ast$}
\]
the cost of the classes of~$D_p$ outside~$\Phi$, and (F1) adds a cheapest
such~$D_p$ (Lemma~\ref{lem:dp}).

\begin{lemma}[Invariants]
\label{lem:invariants}
After every sequence of steps from $(\mathcal{H},\emptyset)$:
\textup{(a)}~the chosen blocks and the pending parts do not overlap:
the sets $\mathit{out}(D_\sigma)$ and $\mathit{Out}(p)$, $p\in\Pi$, are
pairwise disjoint, and the blocks of~$D_\sigma$ have distinct output groups;
\textup{(b)}~$\Phi$ records the paid shareable classes that may still
be used: its classes are classes of~$D_\sigma$, and it contains every
shareable class of~$D_\sigma$ that a pending group can use;
\textup{(c)}~$J(\sigma)$ is the cost of the chosen blocks, each class
counted once: $J(\sigma)=\sum_{c\in\classes(D_\sigma)}w_\theta(c)$;
\textup{(d)}~every completion is a decomposition:
$D_\sigma\cup\bigcup_{p\in\Pi}D_p\in\Dspace$ for every choice of
$D_p\in\mathcal{D}(p)$, $p\in\Pi$.
In particular, at a goal~$\sigma$, $D_\sigma\in\Dspace$ and
$J(\sigma)=J_\theta(D_\sigma)$.
\end{lemma}

\begin{proof}
By induction on the number of steps. Initially, $D_\sigma=\Phi=\emptyset$,
$J(\sigma)=0$, and $\Pi=\mathcal{H}$, so (a) holds by
Lemma~\ref{lem:tree}(a), (b) and (c) hold trivially, and (d) is
Lemma~\ref{lem:tree}(c). We show that every step preserves (a)--(d).

\emph{Structure: (a) and~(d).} A move replaces a pending group~$p$ by
the block~$b$ given to it and the non-forced inputs~$i$ of~$b$; step~(F1)
replaces~$p$ by the blocks of $D^{*}=D^{*}_{w_\Phi}(p)\in\mathcal{D}(p)$. In
both cases, the new blocks and pending groups lie within $\mathit{Out}(p)$,
which was disjoint from $\mathit{out}(D_\sigma)$ and from the sets
$\mathit{Out}(p')$ of the other pending groups~$p'$; within it, $p$ and the
sets $\mathit{Out}(i)$ are pairwise disjoint, and the output groups of~$D^{*}$
are distinct, by Lemma~\ref{lem:tree}(a,b). So (a) still holds. For~(d), a
completion after the step is one before it, with
$D_p=\{b\}\cup\bigcup_{i}D_i$ or $D_p=D^{*}$. Step~(F2) changes neither
$D_\sigma$ nor~$\Pi$.

\emph{Record: (b).} A move or~(F1) only replaces a pending group by its inputs
or removes it, so the pending groups can use no new class. A move adds the
class of its block to the classes of~$D_\sigma$, and to~$\Phi$ if the class is
shareable; the shareable classes of~$D^{*}$ in~(F1) can be used by~$p$, so
they already lie in~$\Phi$; and (F2) drops only classes that no pending group
can use. So (b) still holds.

\emph{Charge: (c).} Let a move give~$p$ a block~$b$ of class~$c$. It pays
$w_\theta(c)$ if $c\notin\Phi$ and nothing otherwise, so it suffices that
$c\in\Phi$ exactly when $c$ is already a class of~$D_\sigma$. If $c\in\Phi$,
it is one by~(b). Conversely, if $c$ is a class of~$D_\sigma$, it is
shareable, as otherwise its only block~$b$ would lie in~$D_\sigma$, and
$p\in\mathit{out}(D_\sigma)\cap\mathit{Out}(p)$ would contradict~(a); and $p$
can use~$c$, so $c\in\Phi$ by~(b). Step~(F1) pays $w_\Phi(D^{*})$, which
by~($\ast$) is the cost of the classes of~$D^{*}$ outside~$\Phi$; the classes
in~$\Phi$ are already paid, and the others are new, as their only blocks lie
in~$D^{*}$, whose output groups are not in $\mathit{out}(D_\sigma)$ by~(a).
Step~(F2) pays nothing and changes no class of~$D_\sigma$.

At a goal, $\Pi=\emptyset$, so (d) gives $D_\sigma\in\Dspace$, and (c) gives
$J(\sigma)=J_\theta(D_\sigma)$.
\end{proof}

\begin{lemma}[Optimality preservation]
\label{lem:complete}
For every $D\in\Dspace$, the search graph has a path from the start state to
a goal~$\sigma$ with $J(\sigma)\le J_\theta(D)$.
\end{lemma}

\noindent So \textsc{Fold} may discard decompositions, but never the
optimal cost: every decomposition has a reachable goal of no greater cost.

\begin{proof}
Fix $D\in\Dspace$. Along the steps from $(\mathcal{H},\emptyset)$, we
keep a decomposition
\[
E=D_\sigma\cup\textstyle\bigcup_{p\in\Pi}E_p,\qquad E_p\in\mathcal{D}(p),
\]
that completes the current state, with $J_\theta(E)\le J_\theta(D)$.
Initially, $E=D$, split as in Lemma~\ref{lem:tree}(c). Each move gives the
first pending group~$p$ the block~$b$ of~$E_p$ with output~$p$; then
$E_p=\{b\}\cup\bigcup_{i}E_i$ with $E_i\in\mathcal{D}(i)$ for the non-forced
inputs~$i$ of~$b$, and $E$ does not change. Step~(F2) does not change~$E$
either. Step~(F1) replaces~$E_p$ by $D^{*}=D^{*}_{w_\Phi}(p)$, and the
result~$E'$ is again a decomposition, by Lemma~\ref{lem:invariants}(d). Only
the classes of~$E_p$ and~$D^{*}$ outside~$\Phi$ can change the cost: those
in~$\Phi$ are classes of~$D_\sigma$ by Lemma~\ref{lem:invariants}(b), and the
others are not shareable and have their only blocks in~$E_p$ or~$D^{*}$,
whose output groups lie in $\mathit{Out}(p)$ and so, by
Lemma~\ref{lem:invariants}(a), nowhere else in~$E$ or~$E'$. Hence, by~($\ast$)
and Lemma~\ref{lem:dp},
\[
J_\theta(E')-J_\theta(E)=w_\Phi(D^{*})-w_\Phi(E_p)\le0.
\]
Each move or (F1) step replaces a pending group by groups below it or
removes it, so, as the memo is finite and acyclic, these steps reach a
goal~$\sigma$, where $E=D_\sigma$. By Lemma~\ref{lem:invariants},
$J(\sigma)=J_\theta(D_\sigma)=J_\theta(E)\le J_\theta(D)$.
\end{proof}

\begin{lemma}[Heuristic]
\label{lem:heuristic}
For every edge from~$\sigma$ to~$\sigma'$ with cost~$k$ between states reached
from the start state, $h(\sigma)\le k+h(\sigma')$, and $h(\sigma)=0$ at every
goal; that is, $h$ is consistent. Hence $h$ is admissible: $h(\sigma)$ is at
most the cost of every path from~$\sigma$ to a goal.
\end{lemma}

\begin{proof}
At a goal, $\Pi=\emptyset$, so $h(\sigma)=0$. We show that no step of an edge
decreases~$h$ by more than it pays; adding up over the steps of the edge then
gives $h(\sigma)\le k+h(\sigma')$.
\emph{(M)}~The move gives the first pending group~$p$ a block~$b$ and pays
$w_\Phi(b)$. As $\hat w$ makes every shareable class free,
$\hat w(b)\le w_\Phi(b)$, and so, by Lemma~\ref{lem:dp},
\[
\mathit{best}_{\hat w}(p)
\le w_\Phi(b)+\textstyle\sum_{i\in I^{\circ}(b)}\mathit{best}_{\hat w}(i).
\]
The move replaces the term $\mathit{best}_{\hat w}(p)$ of~$h$ by the terms
$\mathit{best}_{\hat w}(i)$, $i\in I^{\circ}(b)$, whose groups are distinct
from each other and from the other pending groups by
Lemma~\ref{lem:invariants}(a), so $h$ decreases by at most $w_\Phi(b)$.
\emph{(F1)}~Solving a group~$p$ pays $\mathit{best}_{w_\Phi}(p)$ and
removes the term $\mathit{best}_{\hat w}(p)$ from~$h$. The two are equal: both
are minima over $\mathcal{D}(p)$, and $w_\Phi$ and~$\hat w$ agree on the
blocks of its members, as $p$ can use no shareable class
outside~$\Phi$.
\emph{(F2)}~Neither~$h$ nor the cost changes.
Summing $h(\sigma)\le k+h(\sigma')$ along a path from~$\sigma$ to a goal,
where $h=0$, shows that $h$ is admissible.
\end{proof}

\begin{lemma}[A*]
\label{lem:astar}
Let a finite graph have nonnegative edge costs and a goal reachable from its
start, and let $h$ be consistent and $0$ at goals. Then A*, which keeps only
the cheapest path found to each state, expands every state at most once and
returns a cheapest path from the start to a goal.
\end{lemma}

\noindent This is the standard optimality of A* with a consistent
heuristic~\cite{hart1968formal}; we include a short proof.

\begin{proof}
For an edge from~$u$ to~$v$ of cost~$k(u,v)$, let
$k_h(u,v)=k(u,v)+h(v)-h(u)$, which is nonnegative as $h$ is consistent. A path
from the start to~$v$ of cost~$\ell$ costs $\ell+h(v)-h(\mathit{start})$
under~$k_h$, so A*, which takes from the queue a state of least $\ell+h$, is
Dijkstra's algorithm on the graph with the costs~$k_h$. With nonnegative
costs, Dijkstra's algorithm takes every state from the queue at most once, with
a cheapest path to it, in the order of the costs of these paths. As $h=0$ at
goals, every path to a goal costs $h(\mathit{start})$ less under~$k_h$ than
under~$k$, so the first goal taken from the queue is reached by a path of least
cost among all paths from the start to a goal.
\end{proof}

\noindent As $\Phi\subseteq\classes_{\mathit{sh}}$ in every state, A* thus
expands at most $2^{|\classes_{\mathit{sh}}|}$ states for each set of pending
groups, as Proposition~\ref{prop:cost} states.

\begin{proof}[Proof of Theorem~\ref{thm:astar}]
By Lemma~\ref{lem:heuristic}, $h$ is consistent and admissible. If no class is
shareable, Algorithm~\ref{alg:search} returns the result of DP, which is
optimal by Theorem~\ref{thm:dp}. Otherwise, every goal~$\sigma$
reached from the start state yields a decomposition $D_\sigma\in\Dspace$ with
$J_\theta(D_\sigma)=J(\sigma)$ (Lemma~\ref{lem:invariants}), and every
$D\in\Dspace$ has such a goal with $J(\sigma)\le J_\theta(D)$
(Lemma~\ref{lem:complete}). Hence
\[
\min_{\sigma}J(\sigma)=\min_{D\in\Dspace}J_\theta(D),
\]
where $\sigma$ ranges over the goals reached from the start state; both
minima exist, as $\Dspace\ne\emptyset$. The search graph is finite, as $\Pi$
is a set of groups and $\Phi$ a set of classes, and its edge costs are
nonnegative. Along a path from the start state, $J$ grows by the cost of the
path, so by Lemma~\ref{lem:astar}, A* returns a goal~$\sigma_G$ that
attains the left-hand minimum, and $D_{\sigma_G}$ attains the right-hand one.
\end{proof}

\section{Prompts}
\label{app:prompts}

\lstdefinestyle{promptstyle}{
  basicstyle=\ttfamily\scriptsize,
  columns=fullflexible,
  breaklines=true,
  breakatwhitespace=true,
  breakindent=0pt,
  showstringspaces=false,
  frame=single,
  framesep=4pt,
  rulecolor=\color{gray!60},
  xleftmargin=4pt,
  xrightmargin=4pt,
  aboveskip=\medskipamount,
  belowskip=\medskipamount
}

We list the prompts of the two semantic measures
(Section~\ref{sec:hardness}) and of the validator. Each is a template in
which a name in braces, such as \texttt{\{question\}}, is filled in for every
call. The other prompts
are in the \texttt{prompts} folder of our artifact
repository.\footnote{\url{https://github.com/Inverse0/validation-aware-decomposition}}

\subsection{Coherence}
\label{app:prompt-coherence}

The relations are the cut points to rate, each written as one SQL statement
over the base tables
(Section~\ref{sec:implementation}). For a relation inside a correlated
subquery, \texttt{\{closed\_form\}} adds an explanation of how it is written.

\begin{lstlisting}[style=promptstyle]
Below are intermediate relations computed inside a SQL query, each written as a stand-alone SQL statement over the base tables.{closed_form} You are not told what the query is for.

For each relation, judge whether it makes sense on its own: whether a person who knows the schema could understand it without seeing the rest of the query.

First write "name": a short noun phrase (at most 8 words) for what the relation is, and "row": what one row of it is (for example "one row per customer"). Then rate "coherence" from 0 to 3:
3 = a natural unit with a clear name and a clear row (for example "customers in Europe"; "number of orders per customer").
2 = makes sense, but the name needs a qualification, or some columns are there only for a later step.
1 = describable only as an operation ("rows of A joined with B"): a staging step whose purpose would show only later, or an unclear row (one row per what?).
0 = does not make sense on its own: for example half of a condition, or columns whose meaning depends on something not shown.

Judge the meaning only: not whether the SQL is correct, and not its length or complexity. A small relation can get 3. Do not guess what the query is for.

SCHEMA
{schema_text}

RELATIONS
{relations}

Answer with ONE JSON object and nothing else: {"C1": {"name": "...", "row": "...", "coherence": n}, "C2": {...}, ...} with exactly the labels of the relations above as keys, every one of them.
\end{lstlisting}

\subsection{Relevance}
\label{app:prompt-relevance}

The relations and \texttt{\{closed\_form\}} are as for coherence;
\texttt{\{evidence\}} holds the additional information given with the
question, if any.

\begin{lstlisting}[style=promptstyle]
A SQL query answers the natural-language question below. The relations below the question are computed inside that query, each written as a stand-alone SQL statement over the base tables.{closed_form}

For each relation, judge whether it is a concept the question mentions. Give "match", one of:
"complete" = it computes a concept the question names or asks for (an entity set, a quantity, the answer itself) with every condition the question attaches to that concept.
"narrower" = it computes such a concept with some of the conditions the question attaches to it applied in full and the others still to come: a natural step toward that concept (for example, the question asks about toy orders placed in 2020 and the relation is the orders placed in 2020).
"split" = one condition of the question is applied only in part: one end of a range the question gives, or a match between two things on only some of the attributes the question matches them on (for example, the question pairs each order with a payment by the same customer on the same day, and the relation pairs orders with payments by the same customer on any day), so that its rows only make sense once a later step applies the rest of that condition. A relation that applies some whole conditions and leaves the others to a later step is narrower, not split, also when it joins tables to apply them (for example, the question asks about customers who placed an order and wrote a review, and the relation is the customers who placed an order).
"technical" = it matches no phrase of the question and splits none: a preparatory step (bringing tables together, a lookup).
"unrelated" = no connection to the question.
Also give "phrase": the words of the question that the relation matches ("" for technical or unrelated), and "missing": for narrower or split, what the question says that is not applied, or applied only in part, in the relation ("" otherwise).

Only what the question says about which rows belong counts (its conditions, including a threshold or a top-k): a computation still to come that does not decide which rows belong (an aggregate over them, the columns shown, a total row) does not make a relation narrower. A small relation that matches one phrase of the question is "complete", exactly like the full answer: being closer to the full answer earns nothing. Do not judge whether the SQL is correct.

QUESTION
{question}
{evidence}
SCHEMA
{schema_text}

RELATIONS
{relations}

Answer with ONE JSON object and nothing else: {"C1": {"match": "...", "phrase": "...", "missing": "..."}, "C2": {...}, ...} with exactly the labels of the relations above as keys, every one of them.
\end{lstlisting}

\subsection{Validation}
\label{app:prompt-validation}

\sys{}, \whole{}, and \wholek{} use the same prompt. For a whole query,
\texttt{\{question\}} is the question and \texttt{\{inputs\_text\}} is empty.
For a block, \texttt{\{question\}} is its requirement followed by judging
instructions, and \texttt{\{inputs\_text\}} describes its inputs with their
established facts (Section~\ref{sec:implementation}).
\texttt{\{schema\_legend\}} explains the notation of the schema.

\begin{lstlisting}[style=promptstyle]
You are reviewing a SQL query written to answer a question. Decide whether the query, exactly as written, answers the question correctly against the schema below. Judge the SQL text itself: its joins, filters, grouping, output columns and the semantics of its operators. Do not rewrite it and do not run anything. Minor formatting differences, column aliases and row order do not matter unless the question asks for an order.

QUESTION
{question}
{evidence}
{schema_legend}
{schema_text}
{inputs_text}
SQL
{sql}

Answer with exactly two lines and nothing else:
VERDICT: yes
REASON: <one sentence>
or
VERDICT: no
REASON: <one sentence>
\end{lstlisting}

\section{User Study Materials}
\label{app:userstudy}

Both questions come from the dw split of BEAVER~\cite{chen2026beaver} and read
three of its tables. We renamed the columns to self-explanatory names, so that
participants could focus on the logic of the queries. We also made each
question precise, following BEAVER's gold SQL, so that the question alone
determines whether a candidate is correct.

\lstdefinestyle{studysql}{
  style=sqlstyle,
  basicstyle=\ttfamily\scriptsize,
  frame=single,
  framesep=4pt,
  rulecolor=\color{gray!60},
  xleftmargin=4pt,
  xrightmargin=4pt,
  aboveskip=\medskipamount,
  belowskip=\medskipamount
}

\paragraph{Task 1 (dw\_4704).}
``For each department in the School of Engineering with a total number of
students enrolled in its courses (over all terms) greater than 50, list the
department name, the total number of enrolled students, and for the term
2019FA, the subject with the highest enrollment offered by that department (if
several subjects tie, list each of them), including the subject title and the
number of enrolled students. Present the results ordered by department name. Do
not return any rounded answers.'' Additional information: ``The School of
Engineering is \texttt{school\_name = 'Engineering'}.''
The correct candidate is:
\begin{lstlisting}[style=studysql]
SELECT
  totals.department_name,
  totals.total_enrollment,
  so1.subject_title AS highest_enrolled_subject,
  so1.num_enrolled_students AS highest_enrollment
FROM (
  SELECT
    dep.department_code,
    dep.department_name,
    SUM(so.num_enrolled_students) AS total_enrollment
  FROM department AS dep
  JOIN subject_offering AS so
    ON so.offering_department_code = dep.department_code
  WHERE
    dep.school_name = 'Engineering'
  GROUP BY
    dep.department_code,
    dep.department_name
) AS totals
JOIN subject_offering AS so1
  ON so1.offering_department_code = totals.department_code
JOIN (
  SELECT
    so2.offering_department_code,
    MAX(so2.num_enrolled_students) AS max_enrollment
  FROM subject_offering AS so2
  WHERE
    so2.term_code = '2019FA'
  GROUP BY
    so2.offering_department_code
) AS dept_max
  ON dept_max.offering_department_code = so1.offering_department_code
  AND so1.num_enrolled_students = dept_max.max_enrollment
WHERE
  totals.total_enrollment > 50 AND so1.term_code = '2019FA'
ORDER BY
  totals.department_name
\end{lstlisting}

\paragraph{Task 2 (dw\_3016).}
``For subjects offered in the term 2019FA, show the subject ID, subject title,
number of enrolled students, department name, department full name, and
indicate whether the number of enrolled students is `HIGH\_ENROLLMENT' (more
than 10\% above the average for that department), `LOW\_ENROLLMENT' (more than
10\% below the average for that department), or `AVERAGE', along with the
department's average enrollment. A department's average is taken over its
subjects offered in 2019FA. Do not return any rounded answers.''
The correct candidate is:
\begin{lstlisting}[style=studysql]
SELECT
  so.subject_id,
  so.subject_title,
  so.num_enrolled_students,
  dep.department_name,
  dep.department_full_name,
  dept_avg.avg_enrollment,
  CASE
    WHEN so.num_enrolled_students > dept_avg.avg_enrollment * 1.1
    THEN 'HIGH_ENROLLMENT'
    WHEN so.num_enrolled_students < dept_avg.avg_enrollment * 0.9
    THEN 'LOW_ENROLLMENT'
    ELSE 'AVERAGE'
  END AS enrollment_category
FROM subject_offering AS so
JOIN department AS dep
  ON so.offering_department_code = dep.department_code
JOIN (
  SELECT
    dep2.department_name,
    AVG(so2.num_enrolled_students) AS avg_enrollment
  FROM subject_offering AS so2
  JOIN department AS dep2
    ON so2.offering_department_code = dep2.department_code
  WHERE
    so2.term_code = '2019FA'
  GROUP BY
    dep2.department_name
) AS dept_avg
  ON dep.department_name = dept_avg.department_name
WHERE
  so.term_code = '2019FA'
\end{lstlisting}

\paragraph{Incorrect candidates.}
Each task has two incorrect candidates; Table~\ref{tab:userstudy-errors} gives
the error of each.

\begin{table}[h]
\centering
\footnotesize
\caption{The errors of the incorrect candidates.}
\label{tab:userstudy-errors}
\begin{tabularx}{\columnwidth}{@{}c c X@{}}
\toprule
Task & Error & Description \\
\midrule
1 & A & The correlated subquery that takes the department's highest enrollment
    lacks \sqlkw{so2.term\_code = '2019FA'}, so it takes the maximum over all
    terms: a department whose maximum is reached only in another term is
    missing. \\
  & B & \sqlkw{ORDER BY so2.num\_enrolled\_students DESC LIMIT 1} keeps one
    subject per department: of several 2019FA subjects tied for the highest
    enrollment, only one is listed. \\
\midrule
2 & A & The \sqlkw{CASE} tests AVERAGE (\sqlkw{>= dept\_avg.avg\_enrollment * 0.9})
    before HIGH, so no subject is \sqlkw{'HIGH\_ENROLLMENT'}. \\
  & B & \sqlkw{SUM(so2.num\_enrolled\_students) / COUNT(*)} divides integers, so
    the department's average is rounded down. \\
\bottomrule
\end{tabularx}
\end{table}

\section{Measures}
\label{app:metrics}

Let $N_{\mathit{inc}}$ and $N_{\mathit{cor}}$ be the numbers of incorrect and
correct candidates (243 and 62), and $N = N_{\mathit{inc}} + N_{\mathit{cor}}$.
An incorrect candidate without a verdict counts in FN, and a correct one in
neither TN nor FP. The measures are
\begin{align*}
\mathrm{Acc} &= \frac{\mathit{TP}+\mathit{TN}}{N}, &
\mathrm{Prec} &= \frac{\mathit{TP}}{\mathit{TP}+\mathit{FP}},\\
\mathrm{Rec} &= \frac{\mathit{TP}}{N_{\mathit{inc}}}, &
\mathrm{Spec} &= \frac{\mathit{TN}}{N_{\mathit{cor}}},\\
F_1 &= \frac{2\,\mathrm{Prec}\cdot\mathrm{Rec}}{\mathrm{Prec}+\mathrm{Rec}}, &
\mathrm{BA} &= \frac{\mathrm{Rec}+\mathrm{Spec}}{2}.
\end{align*}

\end{document}